\documentclass{cernrep}

\usepackage{texnames}
\usepackage[T1]{fontenc}
\usepackage[bookmarks, colorlinks=true, linktoc=page, pdftex, linkcolor=red, citecolor=red, urlcolor=red]{hyperref}
\begin{document}
\title{Neutrinos: Fast \& Curious}
\author{Gabriela Barenboim}
\institute{Departament de F\'isica Te\`orica and IFIC, Universitat de
Val\`encia- CSIC\\  
Carrer Dr. Moliner 50, E-46100 Burjassot (Val\`encia), Spain}
\maketitle

\begin{abstract}
The Standard Model has been effective way beyond expectations in foreseeing the result of almost all the experimental tests done up so far. In it, neutrinos are massless. Nonetheless, in recent years we have collected solid proofs indicating 
little but non zero masses for the 
neutrinos (when contrasted with those of the  charged leptons). 
These masses permit 
neutrinos to change their flavor and oscillate, indeed a unique treat. In these lectures, I discuss 
the properties and the amazing potential of neutrinos in and beyond the Standard Model.
\end{abstract}

\section{Introduction}

Last decade witnessed a  brutal transformation in neutrino physics. 
It has been experimentally observed that neutrinos have nonzero masses, 
implying  that leptons blend. This fact was demonstrated by the 
experimental evidence that neutrinos can change from 
one state, or ``flavour'', to another. 
All the information we have 
accumulated about neutrinos, is quite recent. Less that twenty years old.
Neutrino physics as a solid science is in its teenage years and therefore
as any adolescence, in a wild and very exciting (and excited) state.

However, before jumping into the late "news" about neutrinos, lets understand how and why neutrinos were conceived. 

The '20s saw the death of numerous sacred cows, and physics was no exemption. 
One of physic's most holly principles, energy conservation, apparently 
showed up not to hold inside the subatomic world. 

For some radioactive nuclei, it appeared that a non-negligible fraction 
of its energy simply vanished,  leaving no trace of its presence. 

In 1920, in a (by now famous) letter to a meeting \cite{pauli}, Pauli quasi apologetically wrote,"Dear radioactive Ladies and Gentlemen, ... as a desperate
remedy to save the principle of energy conservation in beta decay, ... I propose the idea of a neutral particle
of spin half". Pauli hypothesised that the missing energy was taken off 
by another particle,  whose 
properties were such that made it invisible and impossible to detect: it had no electric charge, no mass  and only very rarely interacted with matter. 
Along these lines, the neutrino was naturally introduced to the universe of particle physics. 

Before long, Fermi postulated the four-Fermi Hamiltonian in order to describe  beta decay utilising the neutrino, electron, 
neutron and proton. Another field was born: weak interactions took the stage to never leave it. 

Closing the loop, twenty years after Pauli's letter, Cowan and Reines got the experimental signature of anti-neutrinos emitted by a nuclear power plant.

As more particles who participated in weak interactions were found in the years following neutrino discovery, weak interactions got credibility as an authentic new force of nature and the neutrino got to be 
a key element of it. 

Further experimental tests through the span of the following 30 years demonstrated that there were not one but three sort, or ``flavours'' 
of neutrinos (electron neutrinos ($\nu_e $), muon neutrinos ($\nu_\mu $) and tau neutrinos 
($\nu_\tau $)) and that, to the extent we could test, had no mass (and no
 charge) whatsoever. 

The neutrino adventure could have easily finish there, however new analyses in neutrinos coming from the sun  
shown us that the neutrino saga was just beginning.... 

In the canonical Standard Model, neutrinos are completely massless and as a consequence are flavour eigenstates,

\begin{eqnarray}
W^+ \longrightarrow e^+ \; + \; \nu_e   \;\;\;\;\;\; ;\;\;\;\;\;\;   Z  \longrightarrow \nu_e \; + \; \bar{\nu}_e 
 \nonumber \\
W^+ \longrightarrow \mu^+ \; + \; \nu_\mu   \;\;\;\;\;\; ;\;\;\;\;\;\;  Z  \longrightarrow \nu_\mu \; + \; \bar{\nu}_\mu   \\
W^+ \longrightarrow \tau^+ \; + \; \nu_\tau     \;\;\;\;\;\; ;\;\;\;\;\;\; Z  \longrightarrow \nu_\tau \; + \; \bar{\nu}_\tau   
\nonumber
\end{eqnarray}
Precisely because they are  massless, they travel at the speed of light and accordingly their flavour does not change 
from generation up to detection. It is evident then, that as flavour is concerned, zero mass neutrinos are 
not an attractive object to study, specially  when contrasted with quarks.

However, if neutrinos were massive, and these masses where not degenerate (degenerate masses flavour-wise is identical to the zero mass case) would mean that  neutrino mass 
eigenstates exist $\nu_i, i=1,2,\ldots$, each with a mass $m_i$. 
The impact of leptonic mixing becomes apparent 
by looking at the leptonic decays, $W^+ \longrightarrow \nu_i + \overline{\ell_\alpha}$ of the charged vector boson $W$. Where, 
$\alpha = e, \mu$, or $\tau$, and $\ell_e$ refers to the electron, $\ell_\mu$ the muon, or $\ell_\tau$ the tau. 

We  call particle $\ell_\alpha$ as the charged lepton of flavour $\alpha$. 
Mixing basically implies that when 
the charged boson $W^+$ decays to a given kind of charged lepton $\overline{\ell_\alpha}$, the neutrino that goes along is not generally the same mass 
eigenstate $\nu_i$. {\em Any} of the different $\nu_i$ can appear. 

The amplitude for the decay of a vector boson $W^+$ to a particular mix $\overline{\ell_\alpha} + \nu_i$ is given by 
$U^*_{\alpha i}$. The neutrino that is radiated in this decay  alongside the given charged lepton
 $\overline{\ell_\alpha}$ is then
\begin{equation}
|\nu_\alpha > = \sum_i U_{\alpha i}^* \; | \nu_i > ~~ .
\label{eq1}
\end{equation}
This specific mixture of mass eigenstates yields the neutrino 
of flavour $\alpha$.

The different $U_{\alpha i}$ can be gathered in a unitary matrix (in the same way they were collected  in the CKM matrix in the quark sector) that receives the
name of 
the leptonic mixing matrix or $U_{PNMS}$ \cite{Maki:1962mu}. The unitarity of $U$ ensures that each time a neutrino of flavour $\alpha$ 
through its interaction produces a charged lepton, the produced charged lepton 
will always be  $\ell_\alpha$, the charged 
lepton of flavour $\alpha $. That is, a $\nu_e$ produces exclusively
an $e$, a $\nu_\mu$ exclusively a $\mu $, and in a similar way  $\nu_\tau$ a
 $\tau$.

The expression (\ref{eq1}), portraying each  neutrino of a given flavour as a 
linear combination of the three  mass eigenstates, 
can be easily inverted to depict every mass eigenstate $\nu_i$ as an analogous linear combination  of the three flavours:
\begin{equation}
|\nu_i > = \sum_\alpha U_{\alpha i} \;  | \nu_\alpha > ~~ .
\label{eq2}
\end{equation}
The amount of $\alpha$-flavour  (or the $\alpha$-fraction) of $\nu_i$ is
 obviously $|U_{\alpha i}|^2$. 
When a $\nu_i$ interacts  and creates  a charged lepton, this $\alpha$-content (or fraction)  expresses the 
probability that the created charged lepton be of flavour $\alpha$.

\section{Neutrino Oscillations basics}\label{s1.2}

The phenomenon of neutrino morphing, flavour transition or in short oscillation,  can be understood in the following form. 

A neutrino is created or emitted by a source along with a charged lepton 
$\overline{\ell_\alpha}$ of flavour $\alpha$. In this way, at the emission point, the neutrino does have a definite flavour. It is  a $\nu_\alpha$. After that point, after production, the neutrino covers some length (propagates thorough a distance)  $L$ until 
it is absorbed. 

At this point, when it has already covered the distance to the target, the neutrino interacts and these interactions  
create another charged lepton $\ell_\beta$ of flavour $\beta$, which we can detect. In this way, at the target, we can know that  
the neutrino is again a neutrino of definite flavour, a  $\nu_\beta$. 
Of course there is a chance that $\beta \neq \alpha$ 
(for instance, if $\ell_\alpha$ is a $\mu$ however $\ell_\beta$ is a $\tau$), 
then,  all along his journey from the source to the 
identification point, the neutrino has morphed or transformed from a $\nu_\alpha$ into a $\nu_\beta$.

This transition from one flavour to the other, 
$\nu_\alpha \longrightarrow \nu_\beta$, is a 
canonical case of the widely known  quantum-mechanical effect present in a 
variety of two state systems and not a particular property of neutrinos.

Since, as shown clearly  by Eq.~(\ref{eq1}), a $\nu_\alpha$ is truly a 
coherent superposition of the three mass eigenstates 
$\nu_i$, the neutrino that travels since it is born until it is detected, 
can be any of the three $\nu_i$'s. Because of that,  we should include the 
contributions of each of the $\nu_i$ in a coherent way. 
As a consequence, the transition amplitude, Amp($\nu_\alpha \longrightarrow \nu_\beta$) receives a contribution of each $\nu_i$ and turns out to be the product of three pieces. The first factor is the amplitude for the neutrino 
created at the generation point along with a charged lepton $\overline{\ell_\alpha}$ to be, particularly, 
a $\nu_i$. And as we have said already, it is given by $U_{\alpha i}^*$.

The second component of our product is the amplitude for the $\nu_i$ 
made by the source to cover the distance up to the detector . We 
will name this element Prop($\nu_i$) 
for the time being and  will postpone the calculation of its value until later. The last (third) piece is the amplitude  for the charged lepton born out of the
interaction of the neutrino  $\nu_i$ with the target to be, particularly, a $\ell_\beta$. 

Being the Hamiltonian that describes the interactions between neutrinos, charged leptons and charged bosons $W$ bosons hermitian (otherwise probability won't be conserved), it follows that if 
Amp($W \longrightarrow \overline{\ell_\alpha} \nu_i ) = U_{\alpha i}^*$, then Amp
$(\nu_i \longrightarrow \ell_\beta W) = 
U_{\beta i}$. 
In this way, the third and last component of the product the $\nu_i$ contribution is given by $U_{\beta i}$, and
\begin{equation}
\mathrm{Amp}(\nu_\alpha \longrightarrow \nu_\beta) = \sum_i U_{\alpha i}^*  \; \; \mathrm{Prop}(\nu_i) 
\;\; U_{\beta i} ~~ .
\label{eq3}
\end{equation}

\begin{figure}[ht]
\begin{center}
\includegraphics[width=12cm]{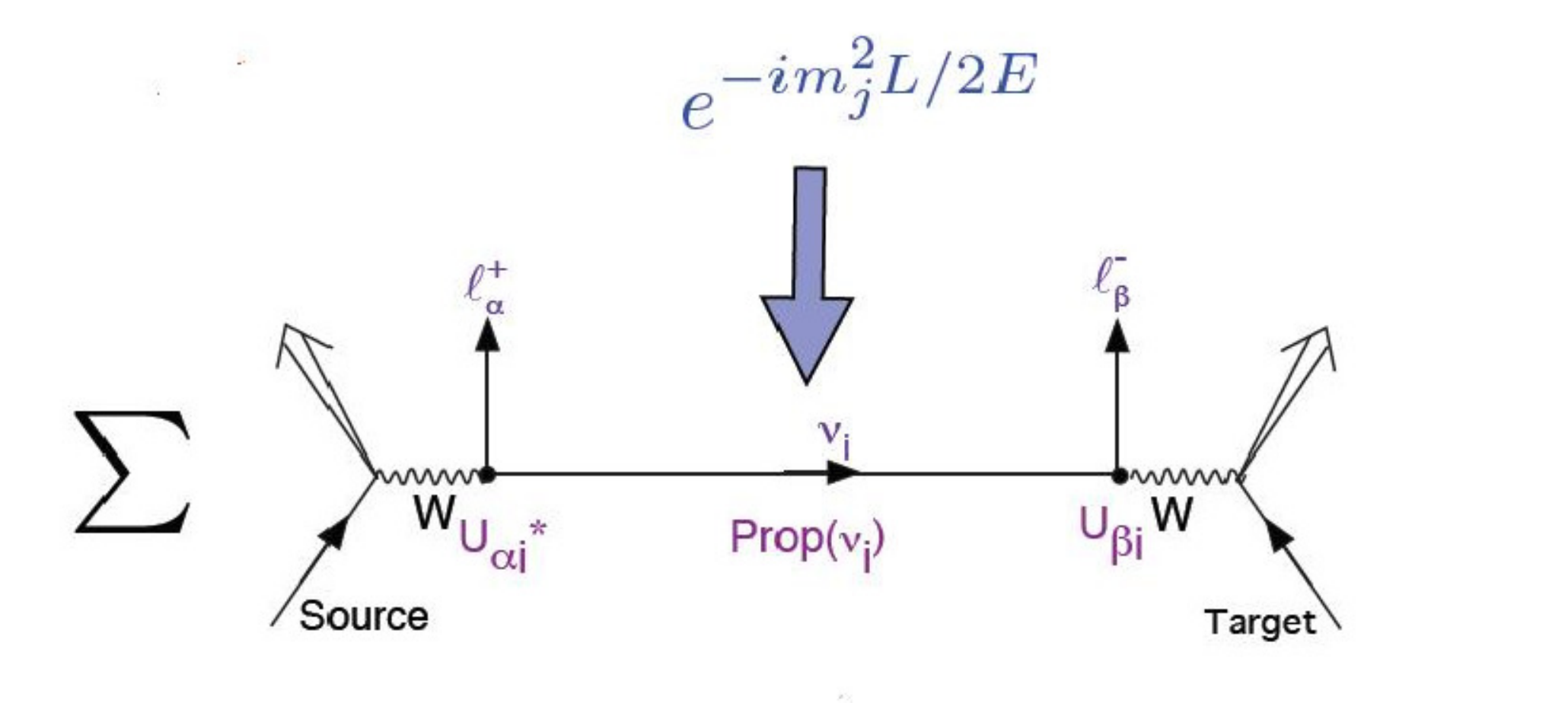}
\caption{Neutrino flavour change (oscillation) in vacuum}
\label{f0}
\end{center}
\end{figure}

It still remains to be established the value  of Prop($\nu_i$). 
To determine it,
we'd better study the $\nu_i$ in its  rest frame. 
We will label the time in that 
system $\tau_i$. If $\nu_i$ does have a rest mass $m_i$, then in this frame of reference its state vector  satisfies the good old Schr\"{o}dinger equation
\begin{equation}
i \frac{\partial}{\partial \tau_i}|\nu_i (\tau_i)>\; = m_i |\nu_i (\tau_i)> ~~ .
\label{eq4}
\end{equation}
whose solution  is given clearly by
\begin{equation}
|\nu_i (\tau_i)>\; = e^{-im_i\tau_i}|\nu_i (0)> ~~ .
\label{eq5}
\end{equation}
Then, the amplitude for a given  mass eigenstate $\nu_i$ to travel freely 
during a time $\tau_i$, 
is simply the amplitude $<\nu_i (0)|\nu_i 
(\tau_i)>$ for observing the initial state  $\nu_i$, $|\nu_i (0)>$  after some time as the 
evoluted  state $|\nu_i (\tau_i)>$, 
{\it ie.} $\exp [-im_i\tau_i]$. Thus Prop($\nu_i$) is only  this amplitude where we have  used that the time taken by
$\nu_i$ to cover the distance from the source to the detector is just $\tau_i$, the proper time.

Nevertheless, if we want Prop($\nu_i$) to be of any use to us, we must write it first 
in terms of  variables we can measure, this means to express it, in variables in the lab frame. 
The natural choice is obviously the distance, $L$, that the neutrino  covers between the
 source and the detector as seen in the lab frame, and the time, $t$, that slips away during the journey, again in the lab frame. The distance $L$ is set by the experimentalists through the
selection of the place of settlement of the source and that of 
the detector and is unique to each experimental setting. Likewise, the value of $t$ is selected by the experimentalists through their election for the time at which the neutrino is made and that when it dies (or gets detected). Therefore, 
$L$ and $t$ are determined (hopefully carefully enough) by the experiment 
design, 
and are the same for all the $\nu_i$ in the beam. 
The different $\nu_i$ do travel through an identical distance $L$, in an identical time $t$.

We still have two additional lab frame variables to determine, the  energy 
$E_i$ 
and three momentum $p_i$ of the neutrino mass  eigenstate $\nu_i$. 
By using the Lorentz invariance of the four component internal product (scalar product), we can obtain  the  expression for the $m_i\tau_i$ appearing in the $ \nu_i$ propagator  Prop($\nu_i$) in terms of the (easy to measure) lab frame variable we have been looking for, which  is given by
\begin{equation}
m_i \tau_i = E_i t - p_i L~~.
\label{eq6}
\end{equation}

At this point however one  may  argue that, in real life, neutrino sources are basically constant in time, and that the time  $t$  that slips away since the neutrino is produced till it dies in the detector is actually not measured. 
This argument is absolutely right. 
In reality, an experiment smears over the time $t$ used by the neutrino to complete its route. However, lets consider that two constituents of the neutrino beam, the first one  with 
energy $E_1$ and the second one  with 
energy $E_2$ (both measured in the lab frame), add up  coherently to the neutrino signal produced in the detector. Now, let us call $t$ to the  the time used  by  the 
neutrino to cover the distance separating the production and detection points. 
Then by the time the constituent whose energy is $E_j \; (j=1,2)$ arrives to the detector, 
it has raised a phase factor $\exp [-iE_j t]$. Therefore, we will have an interference 
between the $E_1$ and $E_2$ beam participants that  will include a phase factor $\exp [-i(E_1 - E_2) t]$. 
When smeared over the non-observed travel time $t$, this factor goes away, {\em except when $E_2 = E_1$}. Therefore, only  
those constituents of the neutrino beam  that share the  same energy  contribute coherently to the neutrino oscillation signal \cite{r4,r3}. Specifically, only the different mass eigenstates constituents of the beams  
that have the same energy weight in. The rest gets averaged out.

Courtesy to is dispersion relation, a mass eigenstate $\nu_i$, with mass $m_i$, and energy $E$, has a three momentum $p_i$ whose absolute value is given by
\begin{equation}
p_i = \sqrt{E^2 - m_i^2} \cong E - \frac{m_i^2}{2E} ~~ .
\label{eq7}
\end{equation}
Where, we have utilised that as the masses of the neutrinos are miserably small, $m_i^2 \ll E^2$ for a typical
energy $E$ attainable at any experiment (the lowest energy neutrinos have MeV energies and sub-eV masses). From Eqs.~(\ref{eq6}) and (\ref{eq7}), it is easy to see that at a given energy $E$ the phase 
$m_i \tau_i$ appearing in Prop($\nu_i$) takes the value
\begin{equation}
m_i \tau_i \cong E(t-L) + \frac{m_i^2}{2E}L ~~ .
\label{eq8}
\end{equation}
As the phase $E(t-L)$ appears in all the interfering terms it will eventually disappear when
calculating the transition amplitude. After all is a common phase factor (its absolute value is one). Thus, we can get rid of it already now and use
\begin{equation}
\mathrm{Prop}(\nu_i) = \exp [-im_i^2 \frac{L}{2E}] ~~ .
\label{eq9}
\end{equation}

Plugging this into Eq.~(\ref{eq3}), we can obtain that the amplitude for a neutrino born as a 
$\nu_\alpha$ to be detected  as a $\nu_\beta$ after covering  a distance $L$ with energy $E$ yields
\begin{equation}
\mathrm{Amp}(\nu_\alpha \longrightarrow \nu_\beta) = \sum_i U_{\alpha i}^* \, e^{-im_i^2 \frac{L}{2E}} U_{\beta i} ~~ .
\label{eq10}
\end{equation}
The expression above is valid  for an arbitrary  number of neutrino flavours and an identical number of  mass eigenstates, as far as they travel through vacuum. The  probability P($\nu_\alpha \longrightarrow \nu_\beta$) for $\nu_\alpha \longrightarrow \nu_\beta$ can be found  by squaring it, giving
\begin{eqnarray}
\mathrm{P}(\nu_\alpha \longrightarrow \nu_\beta) & = & |\mathrm{Amp}(\nu_\alpha \longrightarrow \nu_\beta)|^2 \nonumber \\
	& = & \delta_{\alpha\beta} - 4\sum_{i>j} \Re (U^*_{\alpha i} U_{\beta i} U_{\alpha j}U^*_{\beta j}) 
\sin^2 \left( \Delta m^2_{ij}\frac{L}{4E}\right) \nonumber \\
	 & & + 2\sum_{i>j} \Im (U^*_{\alpha i} U_{\beta i} 
U_{\alpha j}U^*_{\beta j}) \sin\, \left(\Delta m^2_{ij}\frac{L}{2E} \right) ~~ ,
\label{eq11}
\end{eqnarray}
with
\begin{equation}
\Delta m_{ij}^2 \equiv m_i^2 - m_j^2 ~~ .
\label{eq12}
\end{equation}
In order to get Eq.~(\ref{eq11}) we have used that the mixing matrix $U$ is unitary.

The oscillation probability P($\nu_\alpha \longrightarrow \nu_\beta$) we have just obtained 
corresponds to that of a {\em neutrino},  and not to an {\em antineutrino}, as  we have used that the oscillating 
neutrino  was produced  along with a charged {\em antilepton} $\bar{\ell}$, and gives birth to
 a charged {\em lepton} $\ell$ once it reaches the detector. 
The corresponding probability P($\overline{\nu_\alpha} \longrightarrow \overline{\nu_\beta}$) 
for an antineutrino oscillation can be obtained from P($\nu_\alpha \longrightarrow \nu_\beta$) taking advantage of  the 
fact that the two transitions $\overline{\nu_\alpha} \longrightarrow \overline{\nu_\beta}$ and $\nu_\beta 
\longrightarrow \nu_\alpha$ are CPT conjugated processes. 
Thus, assuming that neutrino interactions respect  CPT\cite{r3b},
\begin{equation}
\mathrm{P}(\overline{\nu_\alpha} \longrightarrow \overline{\nu_\beta}) = \mathrm{P}(\nu_\beta \longrightarrow \nu_\alpha) ~~ .
\label{eq13}
\end{equation}
Then, from Eq.~(\ref{eq11}) we obtain that 
\begin{equation}
\mathrm{P}(\nu_\beta \longrightarrow \nu_\alpha;\: U) = \mathrm{P}(\nu_\alpha \longrightarrow \nu_\beta; \: U^*) ~~ .
\label{eq14}
\end{equation}
Therefore, if CPT is a good symmetry (as far as  neutrino interactions are concerned), Eq.~(\ref{eq11}) tells us  that
\begin{eqnarray}
\mathrm{P}( \shortstack{{\tiny (\rule[.4ex]{1em}{.1mm})}
  \\ [-.7ex] $\nu_\alpha $}  \longrightarrow 
\shortstack{{\tiny (\rule[.4ex]{1em}{.1mm})}
  \\ [-.7ex] $\nu_\beta $}) 
 & = & \delta_{\alpha\beta} - 4\sum_{i>j} 
\Re (U^*_{\alpha i} U_{\beta i} U_{\alpha j}U^*_{\beta j}) \sin^2 \left( 
\Delta m^2_{ij}\frac{L}{4E} \right) \nonumber \\
	 &  &   + (-) 
\; 2\sum_{i>j} \Im (U^*_{\alpha i} U_{\beta i} 
U_{\alpha j}U^*_{\beta j}) \sin\, \left( \Delta m^2_{ij}\frac{L}{2E} \right) ~~ .
\label{eq15}
\end{eqnarray}
These expressions make it clear that if the mixing matrix $U$  is complex, P($\overline{\nu_\alpha} \longrightarrow 
\overline{\nu_\beta}$) and P($\nu_\alpha \longrightarrow \nu_\beta$) will not be identical, in general. As $\overline{\nu_\alpha} 
\longrightarrow \overline{\nu_\beta}$ and $\nu_\alpha \longrightarrow \nu_\beta$ are CP conjugated processes,  P$(\overline{\nu_\alpha} \longrightarrow \overline{\nu_\beta}) \neq \mathrm{P}(\nu_\alpha \longrightarrow \nu_\beta)$ would provide  evidence of CP violation in neutrino oscillations (if Nature has chosen its mixing parameters so that the mixing matrix is indeed complex). Until now, CP violation 
has been observed  only in the quark sector, so its measurement 
in neutrino physics would be quite exciting.

So far, we have been working in natural units. A fact that becomes transparent by looking at the dispersion relation Eq.~(\ref{eq8}). If we restore now 
the $\hbar$'s and $c$ factors (we have happily set to one) into the oscillation probability we find that 
\begin{equation}
\sin^2 \left( \Delta m^2_{ij}\frac{L}{4E} \right)  \;\;\longrightarrow\;\;
 \sin^2 \left( \Delta m^2_{ij} c^4 \frac{L}{4\hbar c E} \right)
\end{equation}
Having done that, it is easy and instructive to explore the semi-classical limit, 
$\hbar \longrightarrow 0$. In this limit  the oscillation length goes to zero (the oscillation phase goes to infinity) and the oscillations are averaged to 1/2. The interference pattern is lost.
A similar situation appears if we let the mass difference  $\Delta m^2$ become large. 
This is exactly  what happens in the
quark sector (and the reason why we never study quark oscillations despite knowing that mass eigenstates do not coincide with flavour eigenstates).

In terms of real life units (which are not "natural" units), the oscillation phase  is given by
\begin{equation}
\Delta m^2_{ij}\frac{L}{4E} = 1.27 \, \Delta m^2_{ij}(\mathrm{eV}^2) \frac{L\,(\mathrm{km})}{E\, 
(\mathrm{GeV})} ~~ .
\label{eq16}
\end{equation}
then, since $\sin^2 [1.27 \, \Delta m^2_{ij}(\mathrm{eV}^2) L\,(\mathrm{km})/E\, (\mathrm{GeV})]$ can be
experimentally observed ({\it ie.} not smeared out) only if its argument is in a ballpark around one, an experimental set-up
 with a baseline  $L$ (km) and an energy $E$ (GeV) is 
sensitive to neutrino mass squared differences  $\Delta m^2_{ij}(\mathrm{eV}^2)$ of order $\sim [L\,(\mathrm{km})/E\, (\mathrm{GeV}]^{-1}$. 
For example, an experiment with a baseline of $L \sim 10^4$ km, roughly the 
size of Earth's diameter, and $E \sim 1$ GeV would explore mass differences $\Delta m^2_{ij}$ down to $\sim \!10^{-4}$ eV$^2$. 
This fact makes it clear that neutrino long-baseline experiments  can test even miserably small neutrino mass differences. 
It  does so by exploiting the  quantum mechanical interference between amplitudes whose relative phases are given precisely by these super tiny neutrino mass differences, which can be transformed into
sizeable effects by choosing $L/E$  appropriately.

But let's keep analysing  the oscillation probability 
and see whether we can learn more about neutrino oscillations by studying its expression.

It is clear from  $\mathrm{P}( \shortstack{{\tiny (\rule[.4ex]{1em}{.1mm})}
  \\ [-.7ex] $\nu_\alpha $}  \longrightarrow 
\shortstack{{\tiny (\rule[.4ex]{1em}{.1mm})}
  \\ [-.7ex] $\nu_\beta $}) $ that if  neutrinos have  zero mass, in such a way
 that all $\Delta m^2_{ij} = 0$, then, 
$ \mathrm{P}( \shortstack{{\tiny (\rule[.4ex]{1em}{.1mm})}
  \\ [-.7ex] $\nu_\alpha $}  \longrightarrow 
\shortstack{{\tiny (\rule[.4ex]{1em}{.1mm})}
  \\ [-.7ex] $\nu_\beta $})
 =  \delta_{\alpha\beta}$. Therefore,
 the experimental observation that neutrinos can morph from one flavour to a different one 
indicates that neutrinos are not
only massive but also that their masses are not degenerate. Actually, it was precisely 
this evidence the one that proved beyond any reasonable doubt that neutrinos are massive.

However, every neutrino oscillation seen so far has involved at some point neutrinos that travel through matter. 
But the expression we derived  is valid only for flavour change in vacuum, and does not take into account any interaction 
between the neutrinos and the matter traversed between  their source and their detector. Thus, 
the question remains whether  it may be that some unknown  flavour changing interactions between neutrinos and matter are indeed responsible of  
the observed flavour transitions, 
and not neutrino masses. 
Regarding this question, a couple of things should be said. First, although it is true that the Standard Model 
of elementary particle physics contains only massless neutrinos, it provides an amazingly well corroborated
description of weak interactions, and therefore of all the ways a neutrino interacts.  Such a description does not include flavour change. 
Second, for some of the processes experimentally observed where neutrinos do change flavour, matter effects are expected to be miserably small, and on those cases the evidence points towards a dependence  on $L$ and $E$ in the flavour transition probability through the 
combination $L/E$, as anticipated  by the oscillation hypothesis. 
Modulo  a constant, $L/E$ is precisely the proper 
time that goes by  in the rest frame of the neutrino as it covers a distance $L$ possessing an energy $E$. Therefore, these 
flavour transitions behave as if they were a true progression of the neutrino itself over time, and not a result of an interaction 
with matter.

Now, lets explore the case where the leptonic mixing  were trivial. 
This would imply that in the charged boson decay $W^+ \longrightarrow  \overline{\ell_\alpha} + \nu_i$, 
which as we established  has an amplitude $U_{\alpha i}^*$, 
the emerging charged antilepton $\overline{\ell_\alpha}$ 
of flavour $\alpha$ comes along always  with the {\em same} neutrino mass eigenstate $\nu_i$. That is, 
if $U_{\alpha i}^* \neq 0$, then due to unitarity, $U_{\alpha j}$ becomes zero for all $j \neq i$. Therefore, from Eq.~(\ref{eq15}) it is clear that,
$\mathrm{P}( \shortstack{{\tiny (\rule[.4ex]{1em}{.1mm})}
  \\ [-.7ex] $\nu_\alpha $}  \longrightarrow 
\shortstack{{\tiny (\rule[.4ex]{1em}{.1mm})}
  \\ [-.7ex] $\nu_\beta $})  =  \delta_{\alpha\beta}$. Thus, the observation that neutrinos morph indicates non trivial a mixing matrix.

Then, we are left with basically two ways to  detect neutrino flavour change. 
The first one is to observe, in a beam of neutrinos which 
are all created with the same flavour, say  $\alpha$, some amount of neutrinos of a new flavour $\beta$ that is different 
from the flavour $\alpha$ we started with. This goes under the name of appearance experiments. 
The second way is to start with a 
beam of identical $\nu_\alpha$s, whose flux is either measured or known, and observe that after travelling some distance this flux is depleted. Such experiments are called  disappearance experiments.

As Eq.~(\ref{eq15}) shows, the transition probability  in vacuum does not only depend on $L/E$ but also
oscillates with it. It is because of this fact  that neutrino flavour transitions are named ``neutrino oscillations''.
Now  notice also that neutrino transition probabilities do not depend on the 
individual neutrino masses (or masses squared) but on the squared-mass {\em differences}. Thus, oscillation experiments 
can only measure the neutrino mass squared spectrum. Not its absolute scale. Experiments can test the pattern but cannot determine the distance above zero the whole spectra lies.

It is clear that neutrino transitions cannot modify the total flux in a neutrino beam, but simply
alter its distribution  between the different flavours. Actually, from Eq.~(\ref{eq15}) and the unitarity of the $U$ matrix, it is obvious that
\begin{equation}
\sum_\beta \mathrm{P}( \shortstack{{\tiny (\rule[.4ex]{1em}{.1mm})}
  \\ [-.7ex] $\nu_\alpha $}  \longrightarrow 
\shortstack{{\tiny (\rule[.4ex]{1em}{.1mm})}
  \\ [-.7ex] $\nu_\beta $}) = 1 ~~ ,
\label{eq17}
\end{equation}
where the sum runs over all flavours $\beta$, including the original one $\alpha$. Eq.~(\ref{eq17}) makes it transparent
that the probability that a neutrino morphs its flavour, added to  the probability that it keeps the flavour it had at birth, 
is  one.  
Ergo, flavour transitions do not modify the total flux. Nevertheless, some of the flavours $\beta \neq \alpha$ into 
which a neutrino can  oscillate into may be {\em sterile} flavours; that is, flavours that do not  take part in 
weak interactions and therefore escape  detection. If any of the original (active) 
neutrino flux turns into sterile, then an experiment able to measure the total {\em active} neutrino flux---that is, 
the flux associated to those neutrinos that couple to the weak gauge bosons: 
$\nu_e,\; \nu_\mu$, and $\nu_\tau$--- will observe it to be not exactly the original one, but smaller than it.
In the experiments performed up today, no flux was ever missed.

In the literature, description  of neutrino oscillations normally assume that the different mass eigenstates 
$\nu_i$ that contribute coherently to a beam share the same {\em momentum}, rather than the same {\em energy} 
as we have argued they must have. While the supposition of equal momentum is technically wrong, it is an inoffensive mistake, since, as can easily be shown, it conveys to the same oscillation probabilities as the ones we 
have obtained.

A relevant and interesting  case of the (not that simple)  formula for P$(\overline{\nu_\alpha} \longrightarrow \overline{\nu_\beta})$ 
is the case where only two flavours participate in the oscillation. The only-two-neutrino scenario 
is a rather rigorous  
description of a vast number of experiments. In fact only recently (and in few experiments)  a more sophisticated (three neutrino description) was needed to fit observations.
Lets assume then, that only two mass eigenstates, which we will name 
$\nu_1$ and $\nu_2$, and two reciprocal flavour states, which we will name $\nu_\mu$ and $\nu_\tau$, are 
relevant, in such a way that only one squared-mass difference, $m^2_2 - m^2_1 \equiv \Delta m^2$ arises. Even more, 
neglecting phase factors that can be proven to have no impact on oscillation probabilities, the mixing matrix $U$ can be written as
\begin{equation}
 \left( \begin{array}{c}   \nu_\mu  \\ \nu_\tau  \end{array} \right) = \left(   
	\begin{array}{cc}    \phantom{-}\cos\theta & \sin\theta  \\           
             -\sin\theta & \cos\theta   \end{array} \right) 
 \left( \begin{array}{c}   \nu_1 \\ \nu_2  \end{array} \right) 
\label{eq18}
\end{equation}
The unitary mixing matrix $U$ of Eq.~(\ref{eq18}) is just a 2$\times$2 rotation matrix, and as such , parameterized by a single rotation angle $\theta$ 
which is named (in neutrino physics) as the mixing angle. 
Plugging  the $U$ of Eq.~(\ref{eq18}) and the unique $\Delta m^2$ into the general formula  of the transition probability
 P$(\overline{\nu_\alpha} \longrightarrow \overline{\nu_\beta})$, Eq.~(\ref{eq15}), we can readily see  that, for $\beta \neq \alpha$, 
when only two neutrinos are relevant,
\begin{equation}
\mathrm{P}(\overline{\nu_\alpha} \longrightarrow \overline{\nu_\beta}) = \sin^2 2\theta \sin^2 
\left( \frac{\Delta m^2  \; L}{4E}
\right) ~~ .
\label{eq19}
\end{equation}
Moreover, the survival probability, {\it ie.} the probability that the neutrino remains with the same  flavour 
its was created with is, as expected, one minus 
the probability that it changes flavour.

\section{Neutrino Oscillations in a medium}

When we create a beam of neutrinos on earth through an accelerator and send it up to thousand kilometres away to 
a meet detector, the beam does not move through vacuum, but through matter, earth matter. 
The beam of neutrinos then scatters from the particles it meets along the way. Such
a coherent forward scattering can have a large effect on the transition probabilities. 
We will assume for the time being that neutrino interactions with matter are  flavour conserving, as described by the Standard
 Model, and comment on the possibility of flavour changing interactions later. Then  as there are only two types of weak interactions (mediated by charged and neutral currents) the
would be accordingly only two possibilities for this coherent forward scattering from matter particles to take place.
Charged current mediated weak interactions will occur only if and only if the incoming neutrino is an electron neutrino. As only the  $\nu_e$ can exchange charged boson $W$ with an Earth electron. 
Thus neutrino-electron coherent forward scattering via $W$ exchange opens up an extra source of interaction energy
 $V_W$ suffered exclusively  by electron neutrinos. Obviously, this additional energy being from weak interactions origin has to be proportional to $G_F$, the Fermi coupling constant. In addition, the interaction energy coming from
$\nu_e-e$ scattering grows with  the number of targets, $N_e$, the number of electrons per unit volume (given by the density of the Earth). Putting everything together it is not difficult to see that
\begin{equation}
V_W = + \sqrt{2}\, G_F\, N_e ~~ ,
\label{eq20}
\end{equation}
clearly, this interaction energy affects also antineutrinos (in a opposite way though). It changes sign if we switch the $\nu_e$  by $\overline{\nu_e}$.

The interactions mediated by neutral currents correspond to the case where a neutrino in matter interacts with a matter electron, proton, or neutron by exchanging a 
neutral $Z$ boson. 
According to the Standard Model weak interactions are flavour blind. Every flavour of neutrino enjoys them, and the amplitude for this $Z$ 
exchange is always the same. 
It  also teaches us that, at zero momentum transfer, electrons and protons 
couple to the $Z$ boson with equal strength. The interaction has though, 
opposite sign. Therefore, 
counting on the fact that  the matter  through which our neutrino  moves is electrically neutral (it 
contains equal number of electrons and protons), 
 the contribution of both, electrons and protons to coherent forward neutrino scattering through $Z$ exchange will add up to zero. 
Consequently  only interactions with neutrons  will survive so that, the effect of the  $Z$ exchange contribution to the interaction potential energy $V_Z$ 
reduces exclusively  to that with neutrons and will be proportional to $N_n$, the number density of neutrons. It goes without saying that it will be equal to all flavours.
This time, we find that
\begin{equation}
V_Z = -\frac{\sqrt{2}}{2}\, G_F\, N_n ~~ ,
\label{eq21}
\end{equation}
as was the case before,  for $V_W$, this contribution will flip sign if we replace the neutrinos 
by anti-neutrinos.

But if, as we said, the Standard Model interactions do not change neutrino flavour, neutrino flavour 
transitions  or neutrino oscillations point undoubtedly to  neutrino mass and mixing even when neutrinos are propagating through matter. Unless 
non-Standard-Model flavour changing interactions play a role.

Neutrino propagation in matter is easy to understand when analysed  through the time dependent Schr\"{o}dinger equation in the lab frame
\begin{equation}
i \frac{\partial}{\partial t} |\nu(t)>\; = \mathcal{H} |\nu(t)> ~~ .
\label{eq22}
\end{equation}
where, $|\nu(t)\!>$ is a (three component) neutrino vector state, in which each neutrino flavour corresponds to one component. In the same way, the Hamiltonian $\mathcal{H}$ is a (tree $\times$ three) matrix in flavour space. To make our lives easy, lets analyse the case where only two neutrino flavours are relevant, say $\nu_e$ and $\nu_\mu$. 
Then
\begin{equation}  
|\nu(t)> \;= \left(   \begin{array}{c}
					f_e (t) \\ f_\mu (t)  \end{array} \right) ~~ ,
\label{eq23}
\end{equation}
with $f_i (t)^2$ the amplitude of the neutrino to be a  $\nu_i$ at time $t$.
 This time the Hamiltonian, $\mathcal{H}$, is a 2$\times$2 matrix in neutrino flavour space, \ie  $\nu_e-\nu_\mu$ space.

It will prove to be clarifying to work out the two flavour case in vacuum first, and  add matter effects afterwards. Using Eq.~(\ref{eq1}) to express
 $|\nu_\alpha >$ as a linear combination of mass eigenstates, we can see that the $\nu_\alpha - \nu_\beta$ matrix element 
of the Hamiltonian in vacuum, $\mathcal{H}_{\mathrm{Vac}}$, can be written as
\begin{eqnarray}
 <\nu_\alpha | \mathcal{H}_{\mathrm{Vac}} | \nu_\beta > & = &  <\sum_i U^*_{\alpha i} \nu_i | \mathcal{H}_{\mathrm{Vac}} |\sum_j U^*_{\beta j}\nu_j > \nonumber \\
	& = & \sum_j U_{\alpha j} U^*_{\beta j} \sqrt{p^2 + m_j^2} ~~ .
\label{eq24}
\end{eqnarray}
where we are supposing that the neutrinos belong to a  beam where all its mass components (the mass eigenstates) share the same  definite momentum $p$. As we have already mentioned, despite this supposition being technically wrong, it leads anyway to the right  transition amplitude. 
In the second line of Eq.~(\ref{eq24}), we have used that the neutrinos  $\nu_j$ with  momentum $p$,  the mass eigenstates,  are the asymptotic states of the hamiltonian, $\mathcal{H}_{\mathrm{Vac}}$ for which constitute an orthonormal basis, satisfy
\begin{eqnarray}
\mathcal{H}_{\mathrm{Vac}} | \nu_j> = E_j|\nu_j>
\end{eqnarray}
and have the standard dispersion relation, 
$E_j = \sqrt{p^2 + m_j^2}$.

As we have already mentioned, neutrino oscillations are  the archetype  quantum interference phenomenon, 
where only  the {\em relative} phases of the interfering states play a role. Therefore, only the {\em relative} energies of 
these states, which set their relative phases, are relevant. As a consequence, if it proves to be
convenient (and it will), 
we can feel free to happily remove  from the Hamiltonian $\mathcal{H}$ 
any contribution proportional to the identity matrix $I$. As we have said, this subtraction will 
leave  unaffected the differences between the eigenvalues of $\mathcal{H}$, and therefore will leave
unaffected the prediction
of $\mathcal{H}$ for flavour transitions.

It goes without saying that as in this case only two neutrinos are relevant, there are only two mass eigenstates, $\nu_1$ and $\nu_2$, and only one mass  
splitting $\Delta m^2 \equiv m^2_2 - m^2_1$, and therefore there should be, as before a unitary  $U$ matrix  given by Eq.~(\ref{eq18}) which rotates from one basis to the other. 
Inserting it  into Eq.~(\ref{eq24}), and assuming that our neutrinos have low masses  as compared  to their momenta, \ie  $(p^2 + m^2_j)^{1/2} \cong p + m^2_j/2p$, 
and removing from $\mathcal{H}_{\mathrm{Vac}}$ a term proportional to the the identity matrix (a removal
we know is going to be harmless), we get
\begin{equation}
\mathcal{H}_{\mathrm{Vac}} = \frac{\Delta m^2}{4E} \left(
	\begin{array}{cc}
	-\cos 2\theta & \sin 2\theta  \\
	\phantom{-}\sin 2\theta & \cos 2\theta
	\end{array}		\right) ~~ .
\label{eq25}
\end{equation}
To write this expression, the highly relativistic approximation, which says 
that $p \cong E$ is used. Where $E$ 
is the average energy of the neutrino mass 
eigenstates in our neutrino beam of ultra high momentum $p$.

It is not difficult to corroborate  that the Hamiltonian $\mathcal{H}_{\mathrm{Vac}}$ of Eq.~(\ref{eq25}) for the two neutrino scenario would give an identical oscillation probability , Eq.~(\ref{eq19}), as the one we have already obtained in a different way. An easy way to do it is to analyse
the transition probability for the process $\nu_e \longrightarrow \nu_\mu$. 
From   Eq.~(\ref{eq18}) it is clear that in terms of the mixing angle, the electron neutrino state 
composition is
\begin{equation}
|\nu_e> \; = \phantom{-} |\nu_1> \cos\theta + |\nu_2 > \sin \theta ~~ ,
\label{eq26}
\end{equation}
while that of the muon neutrino is given by
\begin{equation}
|\nu_\mu> \; = -|\nu_1> \sin\theta + |\nu_2 > \cos \theta ~~ .
\label{eq27}
\end{equation}
In the same way, we can also write the eigenvalues of the vacuum hamiltonian $\mathcal{H}_{\mathrm{Vac}}$, \Eq{25}, in terms of the mass squared differences as
\begin{equation}
\lambda_1 = -\frac{\Delta m^2}{4E} ~~ , \;\lambda_2 = +\frac{\Delta m^2}{4E} ~~.
\label{eq28}
\end{equation}
The mass eigenbasis of this Hamiltonian, $|\nu_1>$ and $|\nu_2 >$, can also be written in terms of flavour eigenbasis $|\nu_e>$ and $|\nu_\mu>$ by means of
Eqs.~(\ref{eq26}) and (\ref{eq27}). Therefore, the Schr\"{o}dinger equation of Eq.~(\ref{eq22}), with the identification of $\mathcal{H}$ in this case with $\mathcal{H}_{\mathrm{Vac}}$  tells us that if at time $t=0$ we begin from a $|\nu_e>$, then once some
time $t$ elapses this $|\nu_e>$ will progress into the state given by  
\begin{equation}
|\nu (t)> \;= |\nu_1> e^{+i\frac{\Delta m^2}{4E}t} \cos\theta + |\nu_2 >  e^{-i\frac{\Delta m^2}{4E}t} \sin \theta ~~ .
\label{eq29}
\end{equation}
Thus, the probability P$(\nu_e \longrightarrow \nu_\mu)$ that this evoluted neutrino  be detected as a different flavour $\nu_\mu$, from Eqs.~(\ref{eq27}) and (\ref{eq29}), is given by,
\begin{eqnarray}
\mathrm{P}(\nu_e \longrightarrow \nu_\mu) & = & |<\nu_\mu | \nu(t)>|^2   \nonumber \\
	& = & |\sin\theta\cos\theta (-e^{i\frac{\Delta m^2}{4E}t} + e^{-i\frac{\Delta m^2}{4E}t}) |^2  \nonumber \\
	& = & \sin^2 2\theta \sin^2 \left( \Delta m^2 \frac{L}{4E} \right) ~~ .
\label{eq30}
\end{eqnarray}
Where we have substituted the time $t$ travelled by our highly relativistic state by the distance $L$ it
has covered. The flavour transition or oscillation  probability of Eq.~(\ref{eq30}), as expected, is exactly the same we have found before, Eq.~(\ref{eq19}).

We can now move on to analyse  neutrino propagation in matter. In this case, the 2$\times$2 Hamiltonian representing the propagation in vacuum 
$\mathcal{H}_{\mathrm{Vac}}$ receives the two additional contributions we have discussed before, and becomes $\mathcal{H}_M$, 
which is given by
\begin{equation}
\mathcal{H}_M = \mathcal{H}_{\mathrm{Vac}} + 
	V_W \left( \begin{array}{cc} 1 & 0 \\ 0 & 0  \end{array} \right) +
	V_Z \left( \begin{array}{cc} 1 & 0 \\ 0 & 1  \end{array} \right) ~~.
\label{eq31}
\end{equation}
In the new Hamiltonian, the first additional  contribution corresponds to the interaction potential 
 due to the charged bosons exchange, Eq.~(\ref{eq20}). As this interaction is suffered only by  $\nu_e$, 
this contribution is different from zero only in the $\mathcal{H}_M$(1,1) element  or the $\nu_e - \nu_e$ element. The second additional contribution, the last term   of Eq.~(\ref{eq31}) comes from the  $Z$  boson exchange, 
Eq.~(\ref{eq21}). Since this interaction is flavour blind, it affects every neutrino  flavour in the same
way, its contribution to $\mathcal{H}_M$ is proportional to the identity matrix, and can be safely neglected. Thus
\begin{equation}
\mathcal{H}_M = \mathcal{H}_{\mathrm{Vac}} + 
	\frac{V_W}{2} +
	\frac{V_W}{2}\left( \begin{array}{cc} 1 & 0 \\ 0 & -1 \end{array} \right)~~,
\label{eq32}
\end{equation}
where (for reasons that are going to become clear later) we have divided  the $W$-exchange contribution into two pieces, one proportional to the identity (that we will disregarded in the next step) and, 
a piece that it is not proportional to the identity, that we will keep. Disregarding the first piece as promised, 
we have from  Eqs.~(\ref{eq25}) and (\ref{eq32})
\begin{equation}
\mathcal{H}_M = \frac{\Delta m^2}{4E}  \left( \begin{array}{cc}
	-(\cos 2\theta - A)  &  \sin 2\theta   \\
	 \sin 2\theta  &  (\cos 2\theta - A)  \end{array}  \right)  ~~ ,
\label{eq33}
\end{equation}
where we have defined
\begin{equation}
A\equiv\frac{V_W /2}{\Delta m^2/4E} = \frac{2\sqrt{2} G_F N_e E}{\Delta m^2} ~~.
\label{eq34}
\end{equation}
Clearly, $A$ parameterizes  the relative size of the matter effects as compared to the vacuum contribution given by the neutrino 
squared-mass splitting and signals the situations  when they become important.

Now, if we introduce (a physically meaningful) short-hand notation
\begin{equation}
\Delta m^2_M \equiv \Delta m^2 \sqrt{\sin^2 2\theta + (\cos 2\theta - A)^2}
\label{eq35}
\end{equation}
and
\begin{equation}
\sin^2 2\theta^M \equiv \frac{\sin^2 2\theta}{\sin^2 2\theta + (\cos 2\theta - A)^2} ~~ ,
\label{eq36}
\end{equation}
then the Hamiltonian in a medium $\mathcal{H}_M$ turns out to be 
\begin{equation}
\mathcal{H}_M = \frac{\Delta m^2_M}{4E}  \left( \begin{array}{cc}
	-\cos 2\theta^M  &  \sin 2\theta^M   \\
	 \phantom{-}\sin 2\theta^M  &  \cos 2\theta^M  \end{array}  \right)  ~~ .
\label{eq37}
\end{equation}
and can be diagonalised by inspection, \ie
as a result of our choice, the Hamiltonian  in a medium , $\mathcal{H}_M$, becomes formally indistinguishable
to the  vacuum one, 
$\mathcal{H}_{\mathrm{Vac}}$, Eq.~(\ref{eq25}). 
The difference being that in this case what used to be the vacuum 
parameters $\Delta m^2$ and $\theta$ are presently given by the matter ones, $\Delta m^2_M$ and $\theta^M$, respectively.

Obviously, the mass eigenstates and eigenvalues (which determine the mixing angle)  of  $\mathcal{H}_M$ are not identical to the ones in  vacuum. 
The mass squared difference  of the matter  eigenstates is not the same as  the vacuum  $\Delta m^2$, and the same happens with the mixing angle. The eigenstates in matter, {\it ie.} the files of the unitary matrix that rotates from the flavour  basis to the mass basis, are  different from the vacuum eigenvalues that form the vacuum mixing matrix, and therefore $\theta_M$ is not $\theta$. 
But, the matter Hamiltonian $\mathcal{H}_M$ does indeed contain all 
about the  propagation of neutrinos  in matter, in the same way  $\mathcal{H}_{\mathrm{Vac}}$ contains all about the propagation in vacuum. 

According to Eq.~(\ref{eq37}), $\mathcal{H}_M$ has the same functional dependence  on the matter parameters $\Delta m^2_M$ and $\theta^M$ as the vacuum Hamiltonian $\mathcal{H}_{\mathrm{Vac}}$, Eq.~(\ref{eq25}), on the vacuum ones, $\Delta m^2$ and $\theta$. 
Therefore, $\Delta m^2_M$ can be identified with an effective mass squared difference in matter, 
and accordingly $\theta^M$ can be unidentified with an effective mixing angle in matter.

In a typical experimental set-up where the neutrino beam is generated by an accelerator and sent away  to a detector that is, say, several hundred, or even thousand kilometres away, 
it traverses through earth matter, but only superficially , it does not  get deep into the earth. 
Then, during this voyage the  matter density  encountered  by such a beam can be taken to be approximately constant \footnote{This approximation is clearly not valid for neutrinos that cross the Earth}. 
But if the density of the earth's matter is constant, the same happens with the electron density $N_e$, and the $A$ parameter in which it is incorporated, which after all is determined by it. And it is also true about the  Hamiltonian $\mathcal{H}_M$. They all become approximately constant, and therefore quite identical to the  
vacuum Hamiltonian $\mathcal{H}_{\mathrm{Vac}}$, except for the particular values of their parameters. 
By comparing Eqs.~(\ref{eq37}) and (\ref{eq25}), we can immediately conclude that 
exactly in the same way   $\mathcal{H}_{\mathrm{Vac}}$ gives rise  to 
vacuum oscillations with probability P$(\nu_e \longrightarrow \nu_\mu)$ of Eq.~(\ref{eq30}), 
$\mathcal{H}_M$ must give rise to matter oscillations with  probability 
\begin{equation}
\mathrm{P}_M(\nu_e \longrightarrow \nu_\mu) = \sin^2 2\theta^M \sin^2  \left( \Delta m^2_M \frac{L}{4E} \right) ~~ .
\label{eq38}
\end{equation}
Namely, the transition and survival probabilities in matter are the same as those in vacuum, except that the vacuum parameters  $\Delta m^2$ and $\theta$ 
are now replaced by their matter counterparts, $\Delta m^2_M$ and  $\theta^M$ .

In theory, judging simply by its potential,  matter effects can have very drastic repercussions in the oscillation probabilities.  The exact impact (if any) can be estimated only after the details of the experimental set-up of the experiment in question are given.
As a rule of thumb, to guess the importance of matter effects, we should keep in mind that for neutrinos propagating through the earth's mantle (not deeper than 200 km below the surface) and if the kinematic phase associated to the solar mass difference is still negligible, 
\begin{equation}
A \cong \frac{E}{13 \;\;\mbox{GeV}}
\end{equation}
so that only for beam energies of several GeV matter effects do matter.

And how much do they matter? They matter a lot! From Eq.~(\ref{eq36}) for the matter mixing angle, $\theta^M$, we can appreciate  that even when  the vacuum mixing angle $\theta$ is 
incredible small, say, $\sin^2 2\theta = 
10^{-4}$, if we get to have $A \cong \cos 2\theta$, \ie for energies of a few tens of GeV, then $\sin^2 2\theta^M$ can be brutally enhanced as compared to its vacuum value and can even reach maximal
mixing, {\it ie.}  $\sin^2 2\theta^M = 1$.
 This wild enhancement of a small mixing angle in vacuum up to a sizeable (even maximal) one in matter is the ``resonant'' enhancement, the largest possible version of  the Mikheyev-Smirnov-Wolfenstein effect \cite{r8,r9,r7,r6}. In the beginning of solar 
neutrino experiments, people entertained the idea  that this brutal enhancement  was actually taking place while neutrinos crossed the sun. Nonetheless, as we will see soon  the mixing angle associated with solar neutrinos 
is quite sizeable ($\sim 34^\circ$) already in vacuum \cite{r10}. Then, 
although matter effects on the sun are important and they do enhance the solar mixing angle, unfortunately 
they are not as drastic as we  once dreamt.
Nevertheless, for long-baselines they will play (they are already playing!) a key role in the determination of the ordering of the neutrino spectrum.

\section{Evidence for neutrino oscillations}

\subsection{Atmospheric and Accelerator Neutrinos }

\begin{figure}[ht]
\begin{center}
\includegraphics[width=8cm]{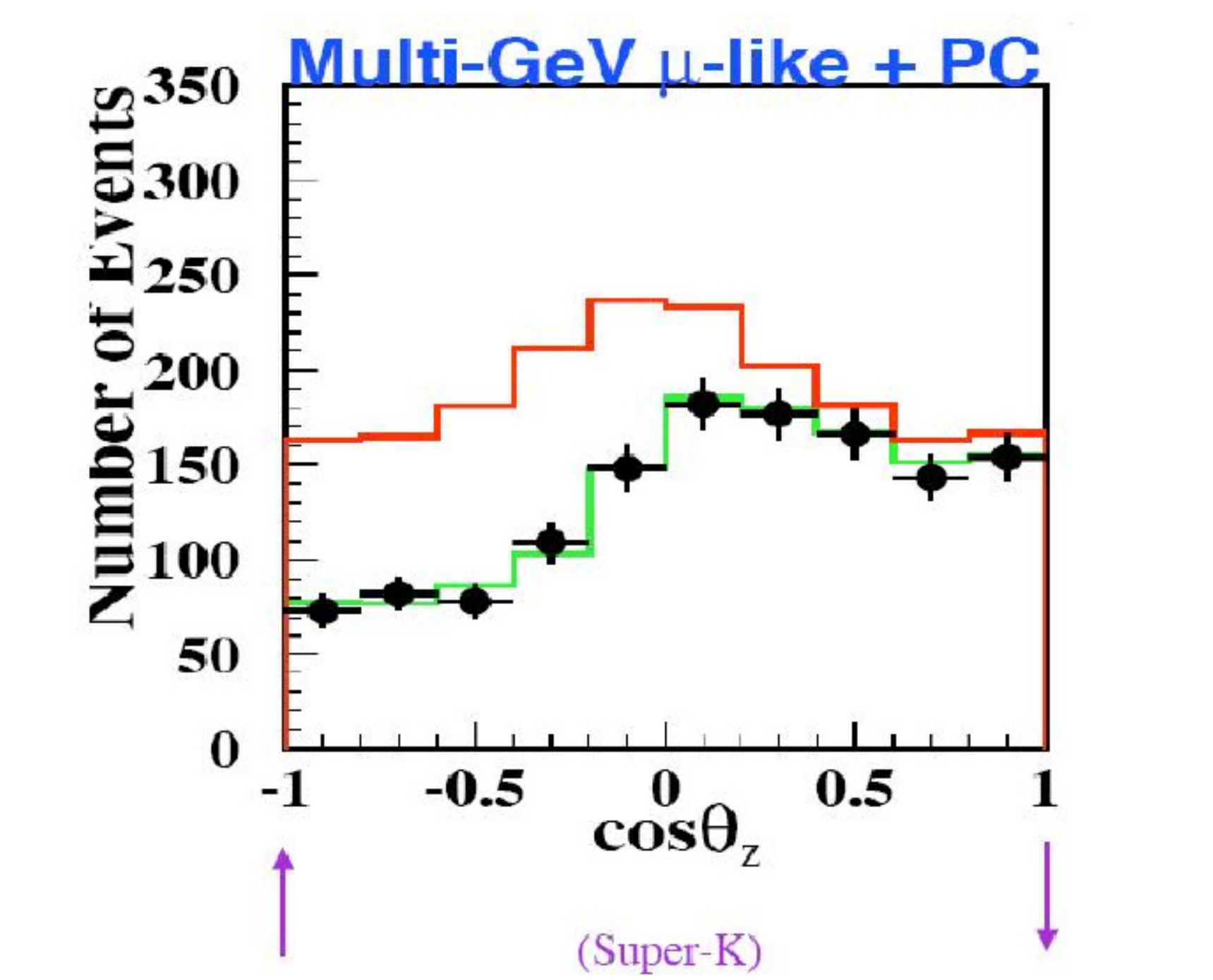}
\includegraphics[width=7.5cm]{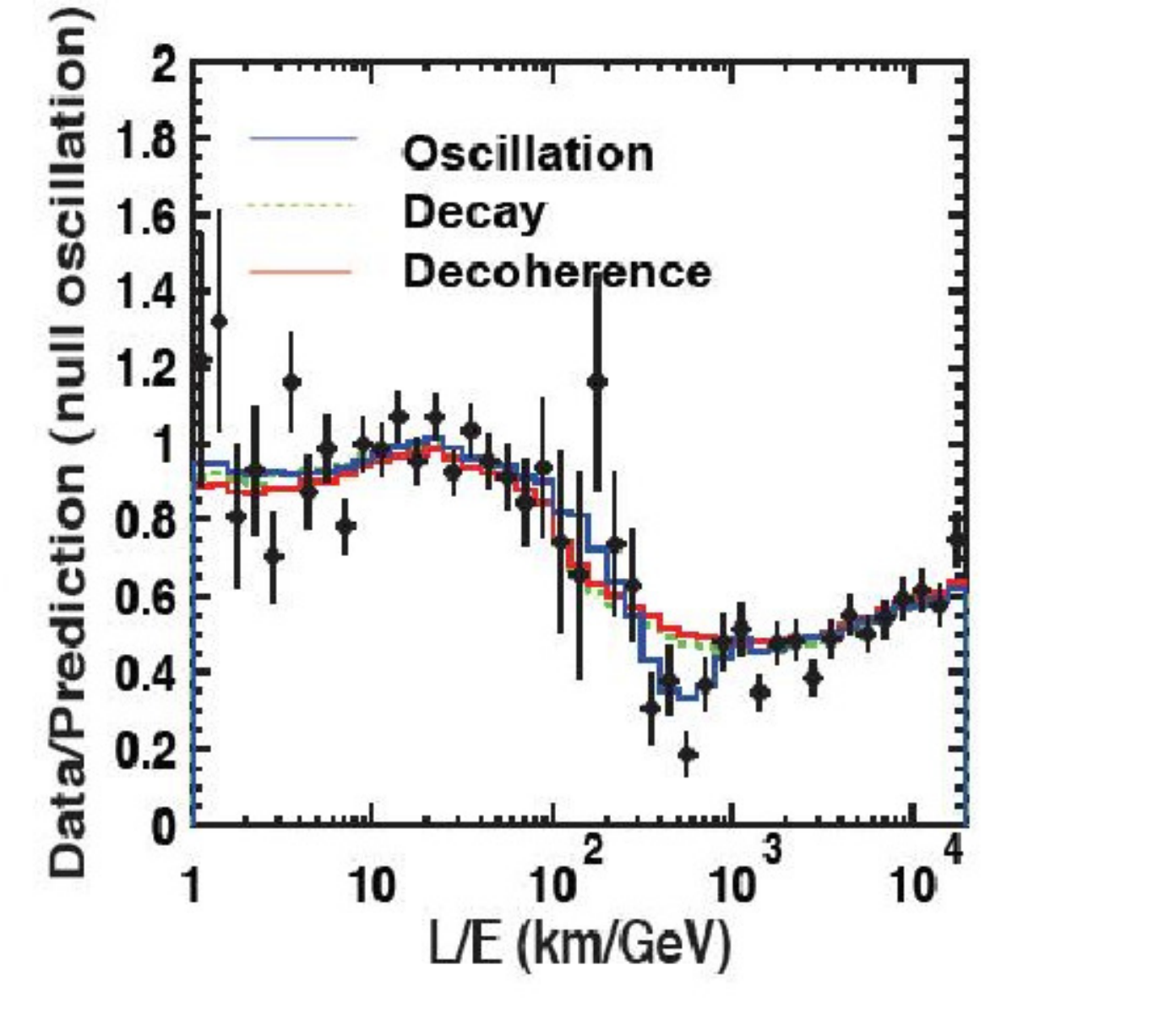}
\caption{Superkamiokande's evidence for neutrino oscillations both in the zenith angle and L/E plots}
\label{f1}
\end{center}
\end{figure}

Almost twenty years have elapsed since we were presented solid and  convincing evidence of neutrino masses and mixings, and since then, the evidence has only grown.
SuperKamiokande (SK) was the first experiment to present 
compelling evidence of $\nu_\mu$ disappearance in their atmospheric neutrino 
fluxes, see \cite{r11} .   In Fig.~\ref{f1} the zenith angle (the angle subtended with the horizontal) dependence of the multi-GeV $\nu_\mu$ sample is shown together 
with the disappearance as a function of $L/E$ plot. These data fit amazingly well  the naive two component neutrino hypothesis with
\begin{equation}
\Delta m^2_{\mbox{atm}} = 2-3 \times 10^{-3} {\mbox{eV}}^2  \;\;\; {\mbox{and}} \;\;\; 
\sin^2 \theta_{\mbox{atm}} = 0.50 \pm 0.13 
\end{equation}
Roughly speaking SK corresponds to  an $L/E$  for oscillations of 500 km/GeV and almost maximal mixing (the mass eigenstates are nearly even admixtures of muon and tau neutrinos).
No signal of an involvement of the third flavour, $\nu_e$ is found so the assumption is that atmospheric neutrino disappearance is basically $\nu_\mu \longrightarrow \nu_\tau$. Notice however, that the first NOvA results seem to point toward a mixing angle which is not maximal (excluding maximal mixing at the 2 sigma level).

\begin{figure}[ht]
\begin{center}
\includegraphics[width=9cm]{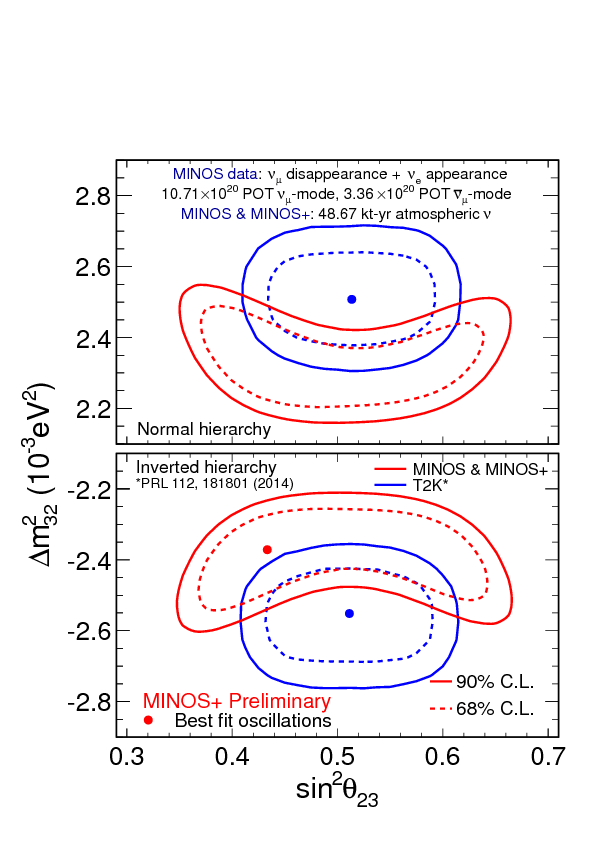}
\caption{Allowed regions in the $\Delta m^2_{\mbox{atm}}$ vs $\sin^2 \theta_{\mbox{atm}}$ plane for MINOS data as well as for
T2K data and two of the SK analyses. MINOS's best fit point is at $\sin^2 \theta_{\mbox{atm}}=.51 $ and
$\Delta m^2_{\mbox{atm}}=  2.37 \times 10^{-3} {\mbox{eV}}^2$. Notice that new NOvA data seem to exclude maximal mixing at the 2 sigma level }
\label{f2}
\end{center}
\end{figure}

After atmospheric neutrino oscillations were established, a new series of neutrino  experiments were built, sending (man-made) beams of $\nu_\mu$ neutrinos to detectors located at large distances:
the K2K (T2K) experiment \cite{r12a,r12b}, sends neutrinos from the  KEK accelerator complex to the old SK mine, with a baseline of 120 (235) km while  the MINOS (NOvA) experiment \cite{r13,r13b}, 
sends its beam from  Fermilab, near Chicago,
to the Soudan mine (Ash river)  in Minnesota, a baseline of 735 (810) km. All these experiments have seen 
evidence for $\nu_\mu$ disappearance consistent with the one found by SK. Their results are summarised in Fig.~\ref{f2}.

\subsection{Reactor and Solar Neutrinos }

The KamLAND reactor experiment, an antineutrino disappearance experiment, receiving neutrinos from sixteen
different reactors, at distances ranging from  hundred to thousand kilometres, with an average baseline of 180 km and neutrinos of a few eV, \cite{r14a,r14b}, has seen evidence of neutrino oscillations . Such
evidence  was collected not only at a  different
$L/E$ than the atmospheric and accelerator experiments but also consists on oscillations
involving electron neutrinos,  $\nu_e$, the ones which were not involved before.
These oscillations have also been seen for neutrinos coming from the sun (the sun produces only electron neutrinos). However,in order to compare the two experiments we should assume that neutrinos (solar) and antineutrinos (reactor) behave in the same way, {\it ie.} assume CPT conservation. The best fit values in the
two  neutrino scenario for the KamLAND experiment are
\begin{equation}
\Delta m^2_\odot = 8.0 \pm 0.4 \times 10^{-5} \mbox{eV}^2  \;\;\; \mbox{and} \;\;\; 
\sin^2 \theta_\odot = 0.31 \pm 0.03 
\end{equation}
In this case, the $L/E$ involved is 15 km/MeV which is more than an order of magnitude larger than the atmospheric scale
and the mixing angle, although large, is clearly not maximal.

\begin{figure}[ht]
\begin{center}
\includegraphics[width=6cm]{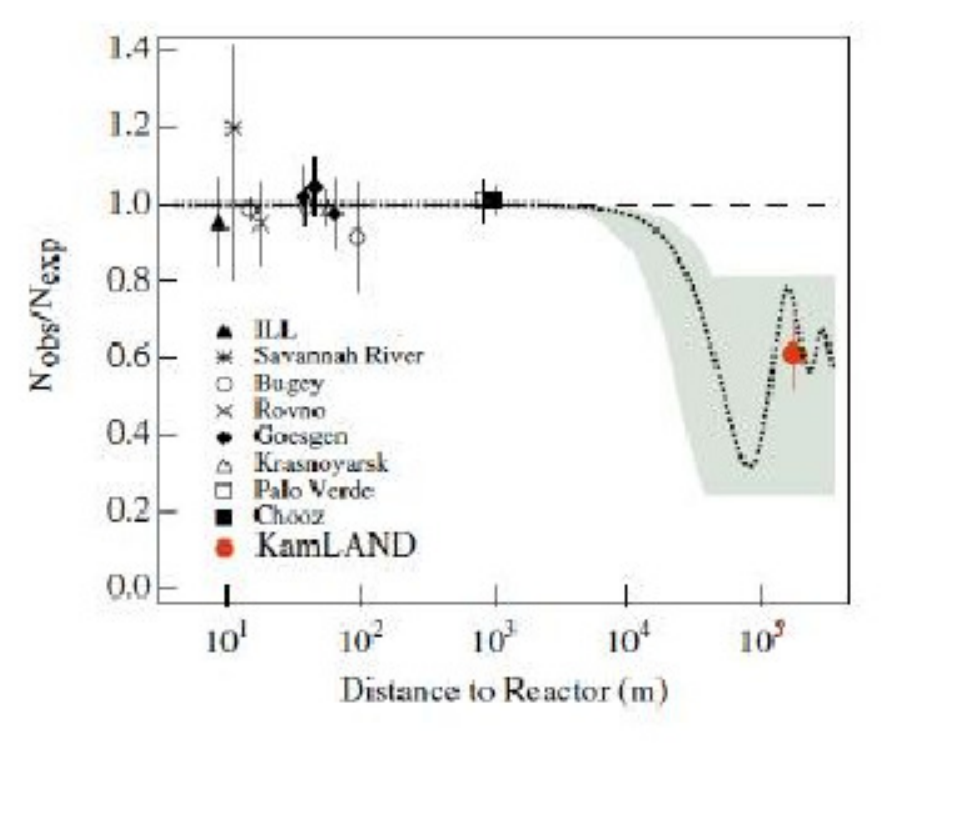}
\includegraphics[width=6cm]{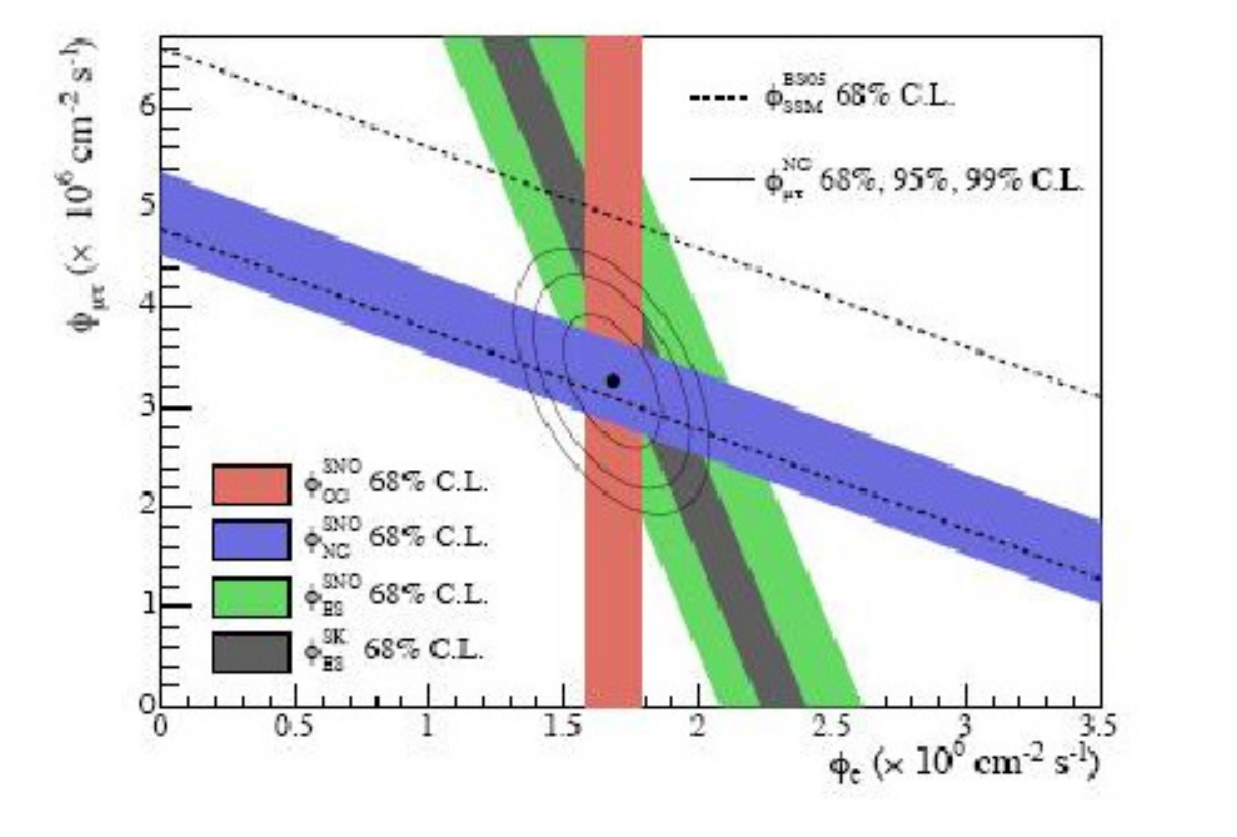}
\caption{Disappearance of the $\bar{\nu}_e $ observed by reactor experiments as a function of distance
from the reactor. The flavour content of the $ ^8$Boron solar neutrinos for the various reactions 
for SNO and SK. CC: $\nu_e + d \longrightarrow e^- + p + p $, NC:$\nu_x + d \longrightarrow \nu_x + p + n $
and ES: $\nu_\alpha + e^- \longrightarrow \nu_\alpha + e^- $ }
\label{f3}
\end{center}
\end{figure}

Fig.~\ref{f3} shows the disappearance probability for the $\bar{\nu}_e$ for KamLAND as well as several older reactor experiments with shorter baselines \footnote{Shorter baseline reactor neutrino experiments, which has seen no evidence of flux depletion suffer the so-called reactor neutrino anomaly, which may point toward the existence of light sterile states}.The second panel 
depicts the flavour content of the $^8$Boron solar neutrino flux (with GeV energies) measured by SNO, \cite{r15}, and SK, 
\cite{r16}.
The reactor outcome can be explained  in terms of two flavour oscillations in vacuum, given that the fit
to the disappearance probability, is appropriately averaged over $E$ and $L$..

The analysis of neutrinos originating from the sun is marginally more complex that the one we did before because it should incorporate  the matter effects that the neutrinos endure since they are born (at the centre of the sun)
until they abandon it, which are imperative at least for the $^8$Boron neutrinos. The pp and 
$^7$Be neutrinos are less energetic and therefore are not significantly altered by the presence
of matter and leave the sun as though it were ethereal. 
$^8$Boron neutrinos  on the other hand, leave the sun unequivocally influenced  by the presence of matter and this is evidenced by the fact that they leave the sun as $\nu_2$, the second  mass eigenstate and 
therefore do not experience oscillations. This distinction among neutrinos coming from different reaction chains is, as mentioned, due mainly  to their 
disparities
at birth. While  pp ($^7$Be) neutrinos are created with  an average energy of 0.2 MeV (0.9 MeV), $^8$B 
are born with 10 MeV and as we have seen the impact of  matter effects grows with the energy of the neutrino.

\begin{figure}[ht]
\begin{center}
\includegraphics[width=7cm]{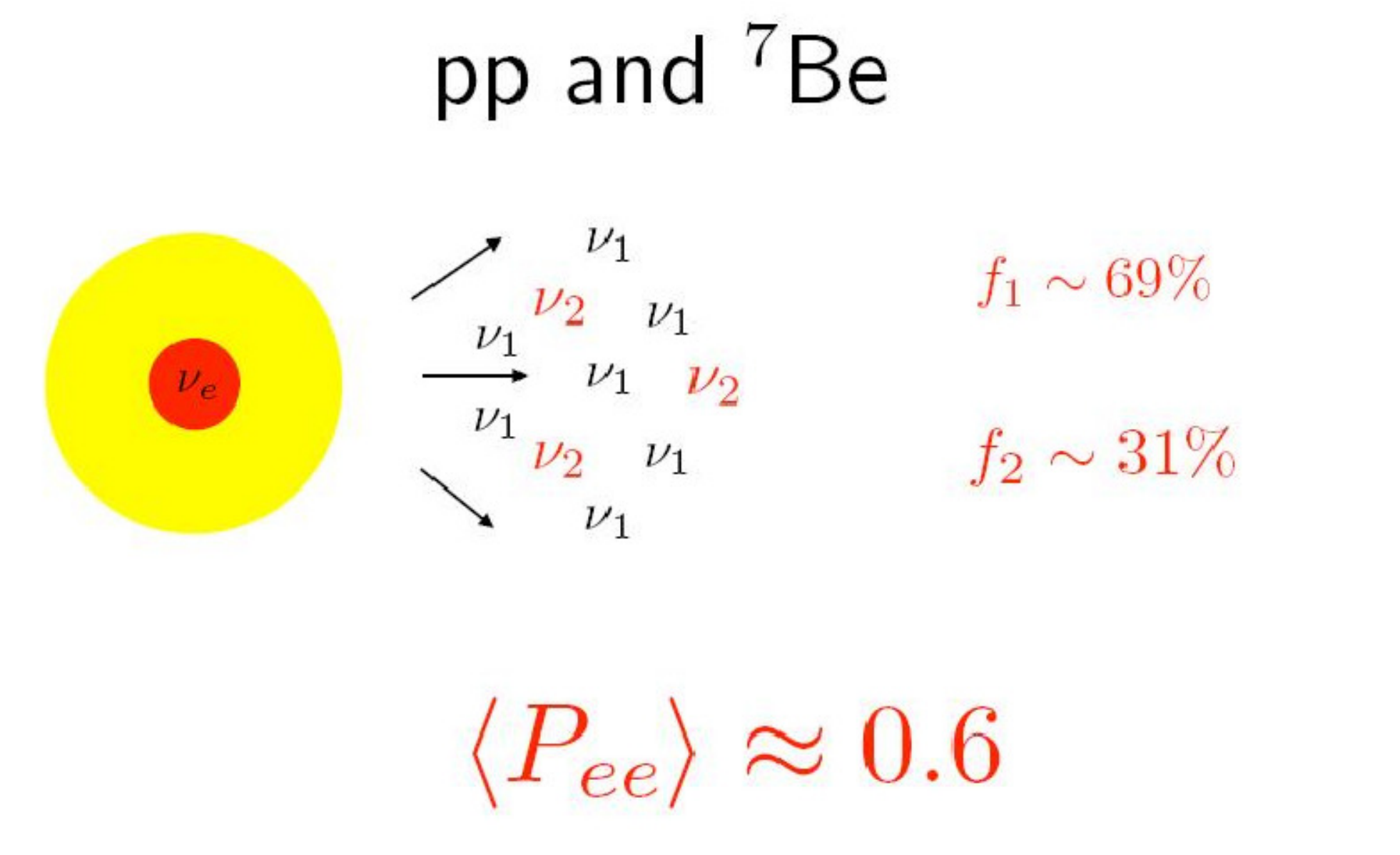}
\includegraphics[width=7cm]{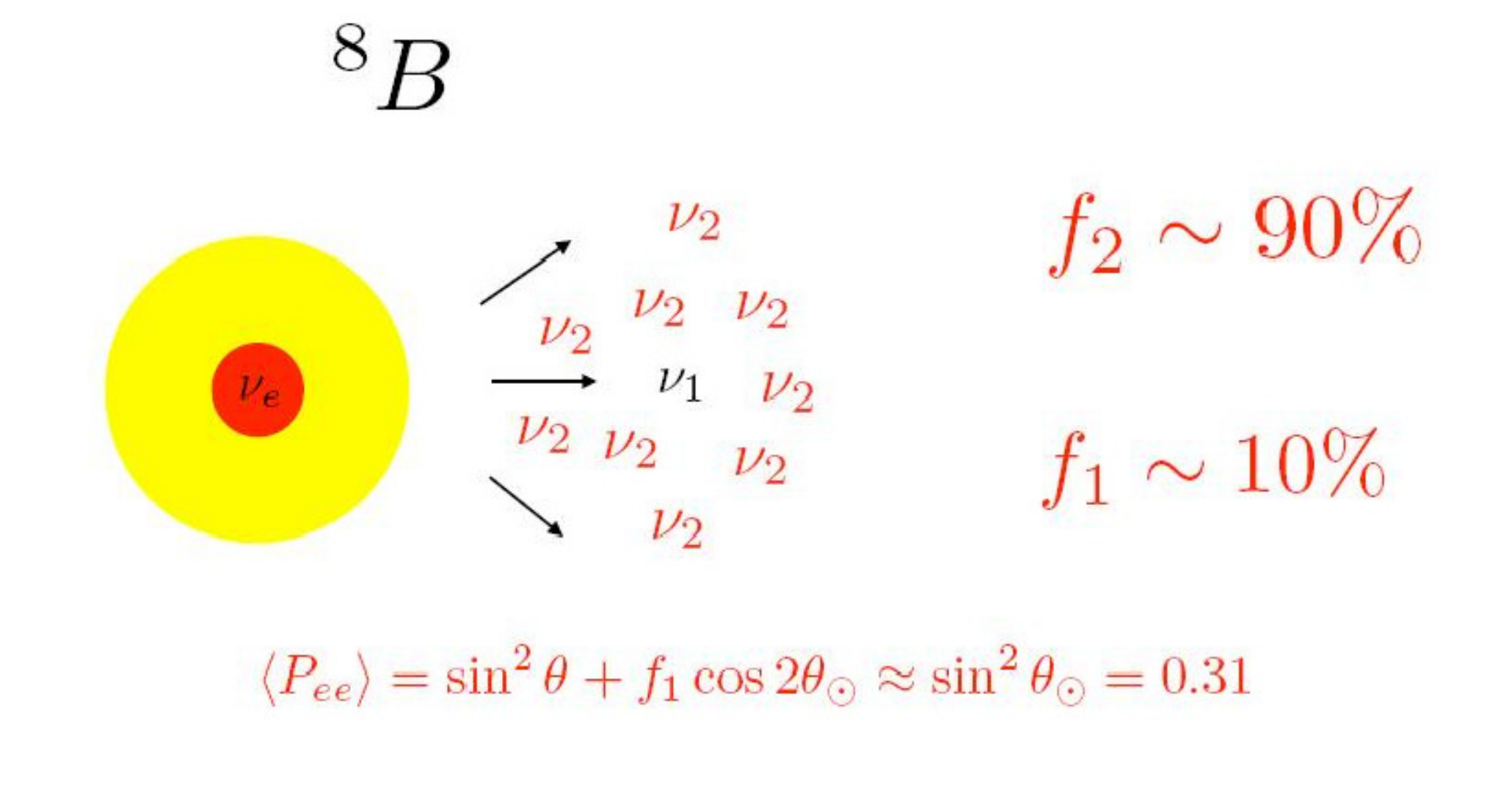}
\caption{The sun produces $\nu_e$ in the core but once they exit the sun thinking about them in the
mass eigenstate basis is useful. The fraction of $\nu_1$ and $\nu_2$ is energy dependent above 1 MeV
and has a dramatic effect on the $^8$Boron solar neutrinos, as first observed by Davis.}
\label{f4}
\end{center}
\end{figure}

However, we ought to emphasise that we do not really see solar neutrino oscillations. To trace the oscillation
pattern, to be able to test is distinctive shape, we need a kinematic phase of order one otherwise the oscillations either do not develop or get averaged to 1/2. In the case of neutrinos coming from the sun the kinematic phase is 
\begin{eqnarray}
\Delta_\odot = \frac{\Delta m^2_\odot L}{4 E} = 10^{7 \pm 1}.
\end{eqnarray}
Consequently, solar neutrinos behave as "effectively incoherent" mass eigenstates once they leave the sun, and remain so once they reach the earth. Consequently the $\nu_e$ 
disappearance or survival probability is given by
\begin{eqnarray}
\langle P_{ee} \rangle = f_1 \cos^2 \theta_\odot + f_2 \sin^2 \theta_\odot
\end{eqnarray}
where $f_1$ is the  $\nu_1$ content or fraction  of $\nu_\mu$ and   $f_2$ is the $\nu_2$ content of $\nu_\mu$
and therefore both fractions satisfy
\begin{eqnarray}
f_1 + f_2 =1.
\end{eqnarray}
Nevertheless, as we have already mentioned, solar neutrinos originating from the pp and $^7$Be chains are not affected by the solar matter and oscillate as in vacuum and thus, in their case $f_1 \approx \cos^2 
\theta_\odot =0.69 $ and  $f_2 \approx \sin^2 \theta_\odot =0.31 $. In the $^8$B a neutrino case, however, 
the impact of solar matter is sizeable and the corresponding fractions are substantially altered, see Fig.~\ref{f4}.

In a two neutrino scenario, the  day-time CC/NC measured by  SNO, which is roughly identical to the day-time average $\nu_e$
survival probability, $\langle P_{ee} \rangle $, reads
\begin{eqnarray}
\left. \frac{CC}{NC}\right|_{\mbox{day}} = \langle P_{ee} \rangle =  f_1 \cos^2 \theta_\odot + f_2 \sin^2 \theta_\odot,
\end{eqnarray}
where $f_1$ and $f_2 = 1 - f_1$ are the $\nu_1$ and $\nu_2$ contents of the muon neutrino, respectively, averaged over
the $^8$B neutrino energy spectrum appropriately weighted with the charged current current cross section. Therefore,
the $\nu_1$ fraction (or how much $f_2$ differs from 100\% ) is given by
\begin{eqnarray}
f_1 = \frac{\left( \left. \frac{CC}{NC}\right|_{\mbox{day}} - \sin^2 \theta_\odot \right)}{\cos 2 \theta_\odot} =
\frac{\left( 0.347 -0.311 \right) }{0.378} \approx 10 \%
\end{eqnarray}
where the central values of the last SNO analysis, \cite{r15}, were used. As there are strong correlations between the 
uncertainties of the CC/NC ratio and $\sin^2 \theta_\odot $ it is not obvious how to estimate the
uncertainty on $f_1$ from their analysis. Note, that if the fraction of $\nu_2$ were 100\%, then 
$ \left. \frac{CC}{NC}\right|_{\mbox{day}} = \sin^2 \theta_\odot $.

Utilising the analytic analysis of the Mikheyev-Smirnov-Wolfenstein (MSW) effect, gave in \cite{r18}, one can obtain the  mass eigenstate 
fractions in a medium, which are  given by
\begin{eqnarray}
f_2 = 1- f_1 = \langle \sin^2 \theta_\odot^M +  P_x \cos 2 \theta_\odot^M \rangle_{^8 {\mbox{B}}},
\end{eqnarray}
with $\theta_\odot^M $ being the mixing angle as given at the $\nu_e$ production point and $P_x$ is the 
probability of the neutrino to hop from one mass eigenstate to the second one  during the  Mikheyev-Smirnov
resonance crossing. The average $\langle ...\rangle_{^8 {\mbox{B}}} $ is over the electron density of the
$^8$B $\nu_e$ production region in the centre of the Sun as given by the Solar Standard Model and the
energy spectrum of $^8$B neutrinos has been appropriately weighted with SNO's charged current cross section.
All in all, the $^8$B energy weighted average content of $\nu_2$'s measured by SNO is
\begin{eqnarray}
f_2 = 91 \pm 2 \% {\mbox{ at the 95 \% C.L.}}.
\end{eqnarray}
Therefore, it is obvious that the $^8$B solar neutrinos are the purest mass eigenstate neutrino beam 
known so far and SK super famous picture of the sun taken (from underground) with neutrinos is made with 
approximately 90\% of $\nu_2$, {\it ie.} almost a pure beam of mass eigenstates.

On March 8, 2012 a newly built reactor neutrino experiment, the Daya Bay experiment, located in China,
announced the measurement of the third mixing angle \cite{An:2012eh}, the only one which was still missing and found it to be
\begin{eqnarray}
\sin^2 (2 \theta_{12}) = 0.092 \pm 0.017 \;
\end{eqnarray}
Following this announcement, several experiments confirmed the finding and during the last years the last mixing angle to be measured became the best (most precisely)  measured one. The fact that this angle, although smaller that the other two, is still sizeable opens the door to a new
generation of neutrino experiments aiming to answer the open questions in the field.

\section{ $\nu $ Standard Model }

Now that we have comprehended the physics behind neutrinos oscillations and have leaned the
experimental evidence about the parameters driving these oscillations, we can move ahead and
construct the  Neutrino Standard Model:
\begin{itemize}
\item it comprises three light ($m_i$  $< $  1 eV) neutrinos, {\it ie.} it involves  just two mass differences

$\Delta m^2_{\mbox{atm}} \approx 2.5 \times 10^{-3} \mbox{eV}^2$ and $\Delta m^2_{\mbox{solar}} \approx 
8.0 \times 10^{-5} {\mbox{eV}}^2$ .

\item so far we have not seen any solid  experimental indication (or need) for additional neutrinos \footnote{Although it must be noted that there are several not significant hint pointing in this direction}. As we have measured long time ago the invisible width of the $Z$ boson and found it to be 3, within errors, 
if additional
neutrinos are going to be incorporated into the model, they cannot couple to the $Z$ boson, {\it ie.} they
cannot enjoy weak interactions, so we call them sterile.  However, as sterile neutrinos have
not been seen (although they may have been hinted), and are not needed to explain any solid experimental evidence, our Neutrino Standard Model will contain just the three
active flavours: $e$, $\mu$ and $\tau$.

\begin{figure}[ht]
\begin{center}
\includegraphics[width=10cm]{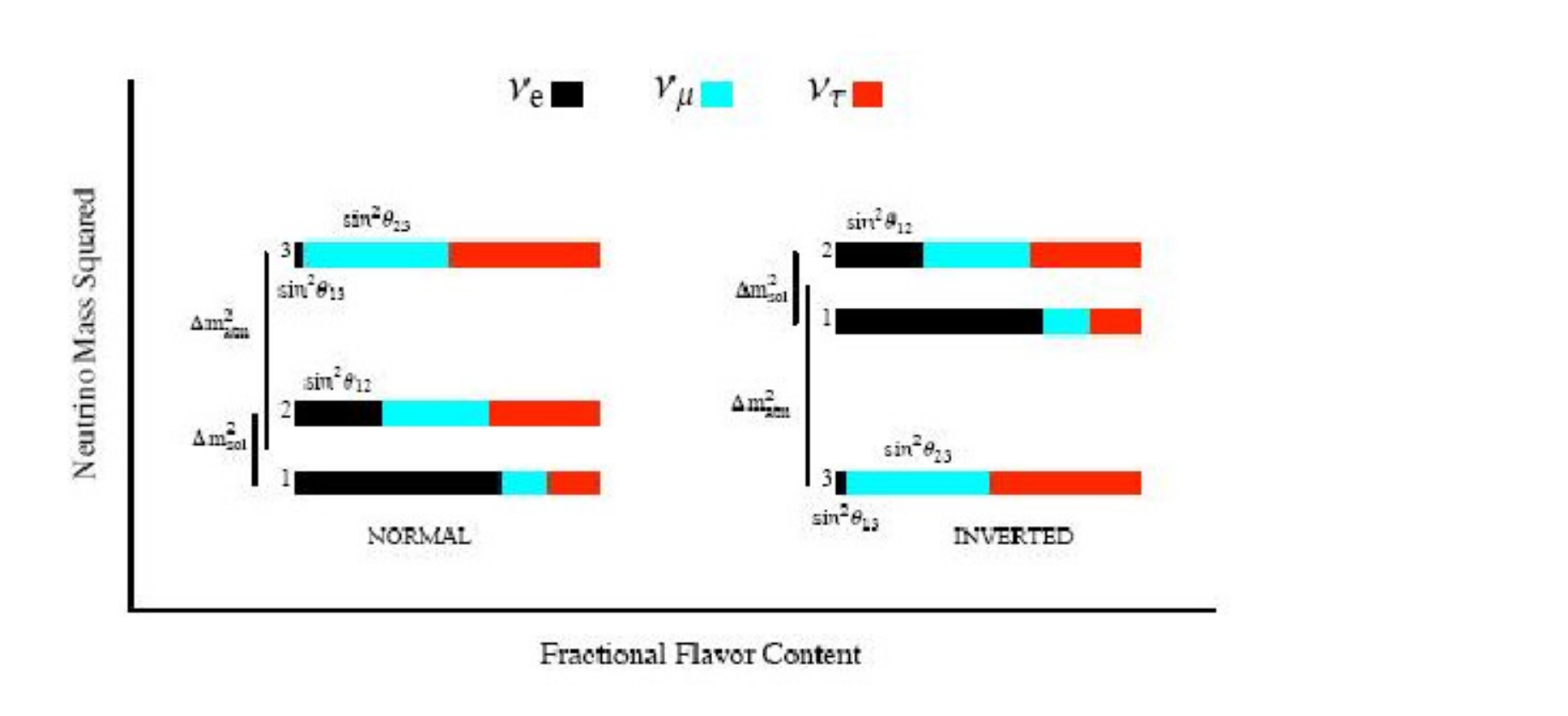}
\caption{Flavour content of the three neutrino mass eigenstates (not including the dependence on the
cosine of the CP violating phase $\delta $).If CPT is conserved, the flavour content must be the same for 
neutrinos and anti-neutrinos. Notice that oscillation experiments cannot tell us how far above zero
the entire spectrum lies.}
\label{f5}
\end{center}
\end{figure}

\item the unitary mixing matrix which rotates from the 
flavour to the mass basis,  called the PMNS matrix,
comprises three mixing angles (the so called solar mixing angle:$\theta_{12}$, the atmospheric mixing angle $\theta_{23}$, and the last to be measured, the reactor mixing angle$\theta_{13}$) , one Dirac phase ($\delta $) and 
potentially two Majorana phases ($\alpha $, $\beta $) and is given by
\begin{eqnarray}
\mid \nu_\alpha \rangle = U_{\alpha i}  \mid \nu_i \rangle \nonumber 
\end{eqnarray}
\begin{eqnarray}
U_{\alpha i } = \left( \begin{array}{ccc} 1 & & \\  & c_{23}& s_{23} \\ & - s_{23}& c_{23} \end{array} \right)
\left( \begin{array}{ccc} c_{13} & & s_{13} e^{-i \delta} \\  & 1& \\ - s_{13} e^{i \delta} &  & c_{13} \end{array} \right)
\left( \begin{array}{ccc}  c_{12}& s_{12} &  \\- s_{12}& c_{12} & \\ &&1 \end{array} \right)
\left( \begin{array}{ccc} 1 & & \\  & e^{i \alpha} & \\ & & e^{i \beta} \end{array} \right)
\nonumber
\end{eqnarray}
where $s_{ij} = \sin \theta_{ij}$ and $c_{ij} = \cos \theta_{ij}$. Courtesy of the hierarchy in mass differences (and to a less extent to the smallness of the reactor mixing angle)  we are permitted  to recognise  the (23) label in the
three neutrino scenario as the
atmospheric $\Delta m^2_{\mbox{atm}}$ we obtained in the two neutrino scenario, in a similar fashion
 the (12) label can be assimilated to the solar $\Delta m^2_\odot $.
The (13) sector drives  the $\nu_e$ flavour oscillations at the atmospheric scale, and the depletion
in reactor neutrino fluxes
see \cite{r19}. 
According to the experiments done so far, the three sigma ranges for the neutrino mixing angles are 
\begin{eqnarray}
0.267 <\; \sin^2 \theta_{12} \; < 0.344 \;\;\;; \;\;\;
0.342 <\; \sin^2 \theta_{23} \;<\; 0.667 \;\;; \;\;\;
0.0156 <\; \sin^2 \theta_{13} \;< 0.0299\nonumber
\end{eqnarray}
while the corresponding ones for  the mass splittings are 
\begin{eqnarray}
2.24  \times 10^{-3} {\mbox{eV}}^2 <\; \mid \Delta m^2_{32} \mid  \; < \;2.70  \times 10^{-3} {\mbox{eV}}^2\nonumber
\end{eqnarray} and
\begin{eqnarray}
 7. \times 10^{-5} {\mbox{eV}}^2 <\;  \Delta m^2_{21} \; < \; 8.09 \times 10^{-5} {\mbox{eV}}^2 . \nonumber
\end{eqnarray}
These mixing angles and mass splittings are summarised in Fig.~\ref{f5}.

\item As oscillation experiments only explore the two mass differences, two ordering are possible, as shown in Fig.~\ref{f5}. They are called normal and inverted hierarchy and roughly identify whether the mass eigenstate with the smaller electron neutrino content is the lightest or the heaviest.

\item The absolute mass scale of the neutrinos, or the mass of the lightest neutrino is
not know yet,  but cosmological bounds already say that the heaviest one must 
be lighter than
about .5 eV. 

\item  As transition or survival probabilities depend on  the combination $U^*_{\alpha i } U_{\beta i}$
no trace of the Majorana phases could appear on oscillation phenomena, however
they will  have observable effects in those processes where the Majorana character of the neutrino
is essential for the process to happen, like neutrino-less double beta decay.

\end{itemize}

\section{Neutrino mass and character}

\subsection{Absolute Neutrino Mass}

The absolute mass scale of the neutrino, {\it ie.} the mass of the lightest/heaviest neutrino,  cannot be obtained from oscillation 
experiments, however this does not mean we have no access to it. 
Direct experiments like tritium beta decay, or neutrinoless double beta decay 
and indirect ones, like cosmological observations, have potential to 
feed us the 
information on the absolute scale of neutrino mass, we so desperately need. 
The Katrin tritium beta decay experiment, \cite{r20}, has
sensitivity down to 200 meV for the "mass" of $\nu_e$ defined as
\begin{eqnarray}
m_{\nu_e} = \mid U_{e1}\mid^2 m_1 + \mid U_{e2}\mid^2 m_2 + \mid U_{e3}\mid^2 m_3.
\end{eqnarray}

\begin{figure}[ht]
\begin{center}
\includegraphics[width=7cm]{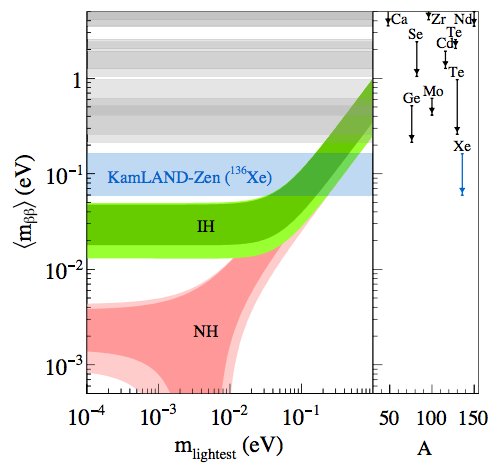}
\includegraphics[width=7cm]{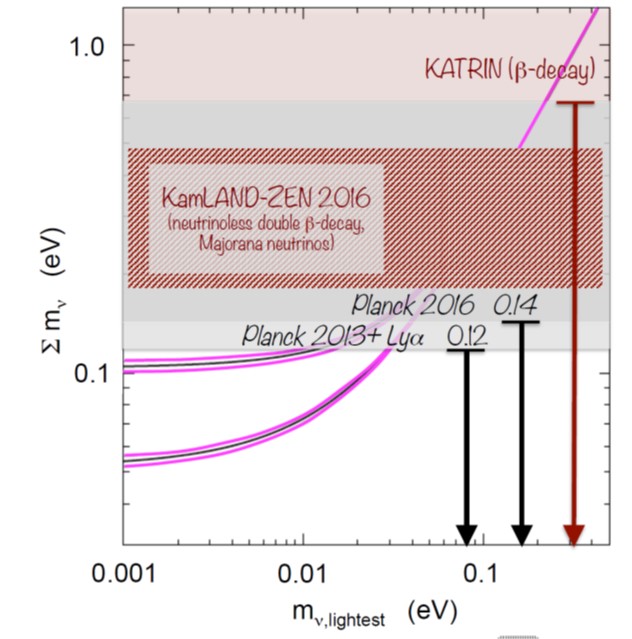}\includegraphics[width=7cm]{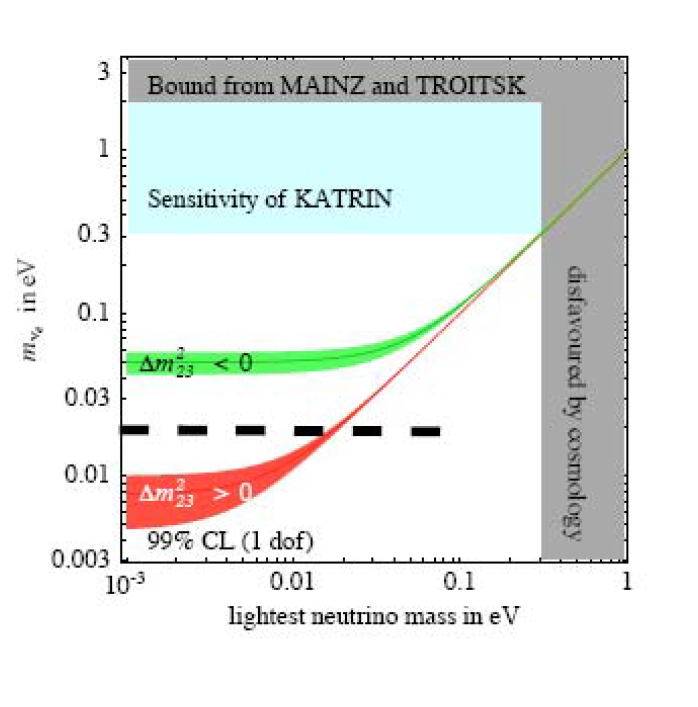}
\caption{The effective mass measured in double $\beta $ decay, in cosmology and in Tritium
$\beta $ decay versus the mass of the lightest neutrino. Below the dashed lines, only the
normal hierarchy is allowed. Notice that while double $\beta$ decay experiments bound the neutrino mass only in the Majorana case, Planck bounds apply for either case}
\label{f6}
\end{center}
\end{figure}

Neutrino-less double beta decay experiments, see \cite{r21} for a review, do not measure the absolute mass of the neutrino directly but a particular combination of neutrino
masses and mixings,
\begin{eqnarray}
m_{\beta \beta} = \mid \sum m_i U_{ei}^2 \mid = \mid m_a c_{13}^2 c_{12}^2 + m_2 c_{13}^2 s_{12}^2 e^{2 i \alpha}
+ m_3 s_{13}^2 e^{2 i \beta} \mid ,
\end{eqnarray}
where it is understood that neutrinos are taken to be Majorana particles, {\it ie.} truly neutral particles (having all their quantum numbers to be zero). 
The new generation
of experiments  seeks to  reach below 10 meV for 
$ m_{\beta \beta} $ in double beta decay.

Cosmological probes (CMB and Large Scale Structure experiments) 
measure the sum of the neutrino masses
\begin{eqnarray}
m_{\mbox{cosmo}} = \sum_{i} m_i .
\end{eqnarray}
and may have a say on the mass ordering (direct or inverted spectrum) as well as test other neutrino properties like neutrino asymmetries\cite{r21b}. If $\sum m_i \approx 10$ eV, the energy balance of the universe saturates the bound coming from its critical density. The current limit, \cite{r22},
 is a 
few \% of this number, $\sim .5$ eV. These bounds are model dependent but they do all give numbers
of the same order of magnitude. However, given the systematic uncertainties characteristic of cosmology, a 
solid limit
of less that 100 meV seems way too aggressive. 

Fig.~\ref{f6} shows the allowed parameter space for the neutrino masses (as a function of the absolute scale) for both the normal and inverted hierarchy.

\subsection{Majorana vs Dirac}

A fermion mass is nothing but a coupling between  a left handed state and a right handed one. Thus, if we examine  a massive fermion at rest, then
one can regard this state as a linear combination of two massless particles, one right handed and one 
left handed. If the particle we are examining is electrically charged, like an electron or a muon, both particles, the left handed  as well as the right handed must have the same charge (we want the mass term to be electrically neutral). This is a Dirac mass term. However, for a neutral particle, like
a sterile neutrino, a new possibility opens up, the left handed particle can be coupled to the right handed
anti-particle, (a term which would have a net charge, if the fields are not absolutely and totally
neutral) this is a Majorana mass term. 

\begin{figure}[!ht]
\begin{center}
\includegraphics[width=7cm]{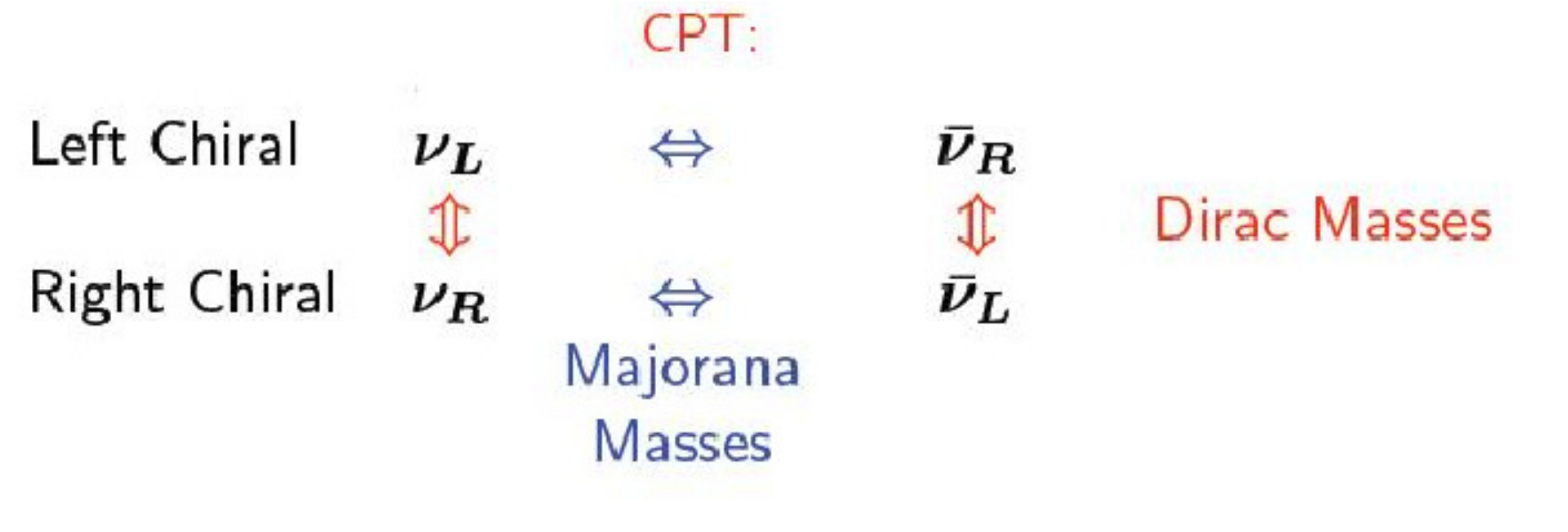}
\label{f7}
\end{center}
\end{figure}

Thus a truly and absolutely neutral particle (who will inevitably be its own antiparticle)  does have two ways of getting a mass term, a la Dirac or a la Majorana, and
if there are no reasons to forbid one of them, will have them both, as
shown in Fig.~\ref{f7}. 

\begin{figure}[!ht]
\begin{center}
\includegraphics[width=7cm]{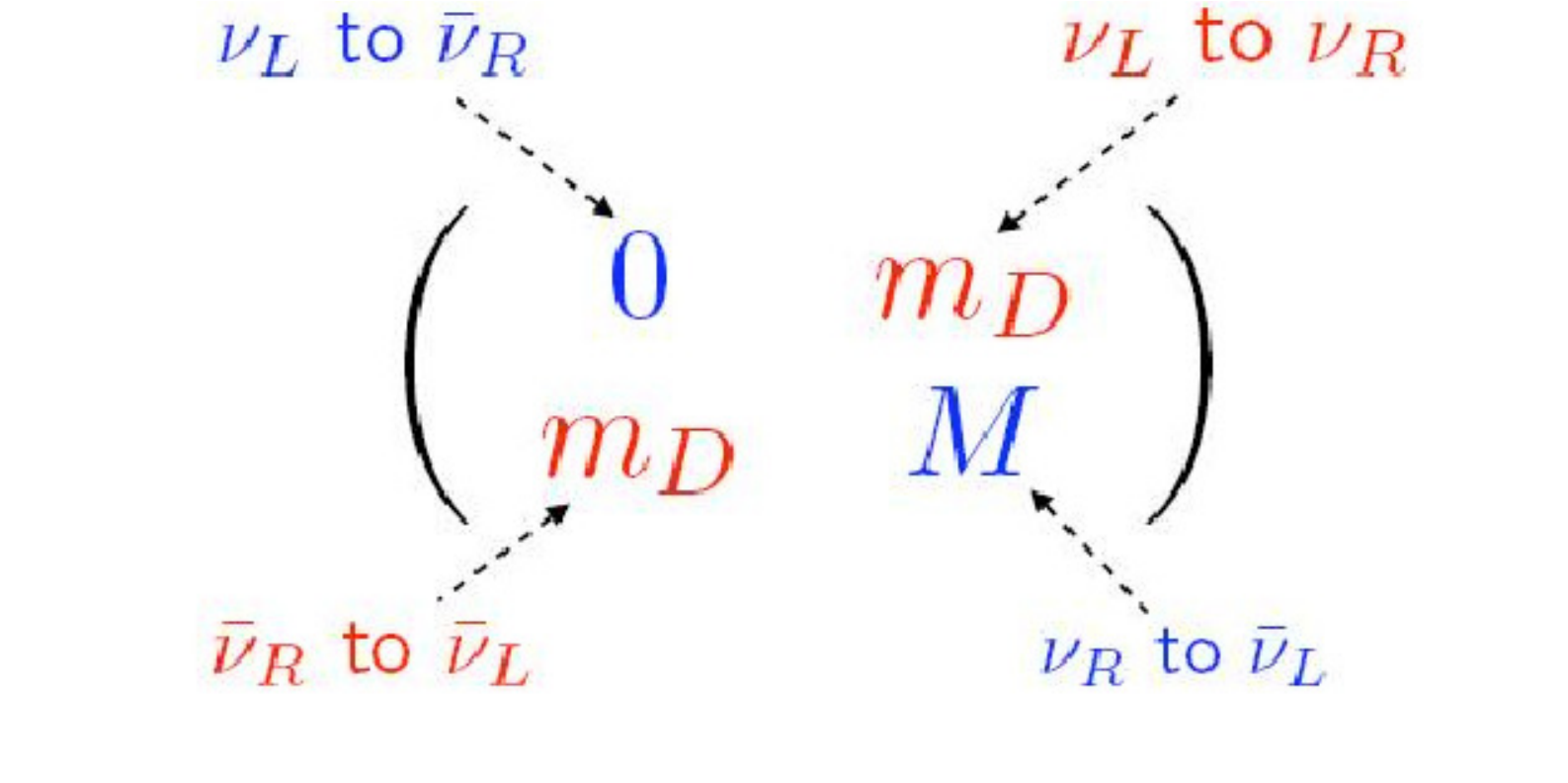}
\caption{The neutrino mass matrix with the various right to left couplings, $M_D$ is the Dirac mass
terms while 0 and $M$ are Majorana masses for the charged and uncharged (under $SU(2) \times U(1)$) chiral
components}
\label{f8}
\end{center}
\end{figure}

In the case of a neutrino, the left chiral field couples to $SU(2) \times U(1) $ implying that a Majorana mass term is
forbidden by gauge symmetry. However, the right chiral field carries no quantum numbers, is totally and absolutely neutral. Then, the
Majorana mass term is unprotected by any symmetry and it is expected to be very large, of the order of the largest scale in the theory. On the other hand, Dirac mass
terms are expected to be of the order of the electroweak scale times a Yukawa coupling, giving a mass
of the order of magnitude of the charged lepton or quark masses. Putting all the pieces together, the mass matrix for the
neutrinos results as in  Fig.~\ref{f8}.

To get the mass eigenstates we need to diagonalise the neutrino mass matrix. By doing so, one is left with two Majorana neutrinos, one super-heavy Majorana
neutrino with mass $\simeq M $ and one super-light Majorana neutrino with mass $m_D^2/M $, {\it ie.} one mass
goes up while the other sinks, this
is what we call the
seesaw mechanism, \cite{r23a,r23b,r23c}\footnote{Depending on the envisioned high energy theory, the simplest see saw mechanism can be categorised into three different classes or types (as they are called) depending on their scalar content.}. The light neutrino(s) is(are) the one(s) observed in current experiments (its mass differences) while the heavy
neutrino(s) are not accessible to current experiments and could be responsible for explaining the
baryon asymmetry of the universe through the generation of a lepton asymmetry at very high energy scales since its decays can in principle be CP violating (they depend on the two Majorana phases on the PNMS
matrix which are invisible for oscillations). The super heavy Majorana neutrinos  being their masses so large can play a role  at very high energies and can be related to inflation \cite{r23d}.

If neutrinos are Majorana particles lepton number is no longer a good quantum number and a plethora of new processes forbidden by lepton number conservation can take place, it is not only 
neutrino-less double beta decay. For example, a muon neutrino can produce  a positively charged muon. However, this
process and any processes of this kind, 
would be suppressed by $(m_\nu / E)^2$ which is tiny, $10^{-20}$, and therefore, although they are 
technically allowed, are experimentally unobservable. To most stringent limit nowadays comes from KamLAND-zen \cite{kamlandzen}, and constraints the half-life  of neutrino-less double beta decay to be $T_{1/2}^{0\nu} > 1.07 \times 10^{26}$ years at 90\% C.L. Forthcoming experiments such as GERDA-PhaseII, Majorana, SuperNEMO, CUORE, and nEXO will improve this sensitivity by one order of magnitude.

Recently low energy sew saw models \cite{low} have experienced a revival and are actively being explored \cite{low2}. In such models the heavy states, of only few tens of TeV can be searched for at the LHC. The heavy right handed states in these models will be produced at LHC either through Yukawa couplings of through gauge coupling to right handed gauge bosons. Some models contain also additional scalar that can be looked for.

\section{Conclusions}

The experimental observations  of neutrino oscillations, meaning that neutrinos have mass and mix, 
answered questions that had endured since the establishment of the Standard Model.
As those veils have disappeared, new questions open up and challenge our understanding
\begin{itemize}
\item  what is the true nature of the neutrinos ? are they Majorana particles or Dirac ones ? are neutrinos totally neutral ? 
\item is there any new scale associated to neutrinos masses ? can it be accessible at colliders ?
\item is the spectrum normal or inverted ? is the lightest neutrino the one with the least electron content on it, or is it the heaviest one ?
\item is CP violated (is $\sin \delta \neq 0 $ ) ? if so, is this phase related at any rate with the baryon asymmetry of the Universe ? what about the other two phases ?
\item which is the absolute mass scale of the neutrinos ? 
\item are there new interactions ? are neutrinos related to the open questions in cosmology, like dark matter and/or dark energy ? do (presumably heavy) neutrinos play a role in inflation ?
\item can neutrinos violate CPT \cite{Barenboim:2002tz}? what about Lorentz invariance ?
\item if we ever measure a different spectrum for neutrinos and antineutrinos (after matter effects are properly taken into account), how can we distinguish whether it is due to a true (genuine) CTP violation or to  a non-standard neutrino interaction ? 
\item are these intriguing signals in short baseline reactor neutrino experiments (the missing fluxes)
a real effect ? Do they imply the existence of sterile neutrinos ? 
\end{itemize}
We would like to answer these questions. For doing it, we are doing right now, and we plan to do
new experiments. These experiments will,
for sure bring some  answers and clearly open new, pressing questions. Only one thing is clear.
Our journey into the neutrino world  is just beginning.

\section*{Acknowledgements}
I would like to thanks the students and the organisers of the European School on HEP for giving me the
opportunity to present these lectures in such a wonderful atmosphere (and for pampering me beyond my expectations, which were not low). I did enjoyed each day of the
school enormously. 
Support from the MEC and FEDER (EC) Grants SEV-2014-0398 and FPA2014-54459 and the Generalitat Valenciana under grant PROMETEOII/2013/017 is acknowledged. The author's work has received funding from the European Union Horizon 2020
research and innovation programme under the Marie Sklodowska-Curie grant
Elusives ITN agreement No 674896  and InvisiblesPlus RISE, agreement No 690575.

\end{document}